\documentclass{aa}
\usepackage{multirow}
\usepackage{comment}
\usepackage[normalem]{ulem}
\usepackage{amsmath}
\usepackage{graphicx,subfig} 
\usepackage{lscape}
\usepackage{longtable}
\usepackage{txfonts}
\usepackage{natbib}
\usepackage{color}
\usepackage{booktabs}
\usepackage{comment}
\usepackage[dvipsnames]{xcolor}
\usepackage{dblfloatfix}

\makeatletter
\renewcommand*\aa@pageof{, page \thepage{} of \pageref*{LastPage}}
\usepackage{hyperref}
\hypersetup{colorlinks=true,allcolors=[rgb]{0,0,0.8}}

\newcommand{\erosita}{eROSITA\xspace}
\newcommand{\planck}{Planck\xspace}
\definecolor{DarkGreen}{HTML}{1D5809}

\newcommand{\esass}{\texttt{eSASS}\xspace}

\newcommand{\erass}[1][1]{eRASS:#1\xspace}

\newcommand{\nh}{\ensuremath{n_{\rm H}}\xspace}
\newcommand{\lcdm}{\ensuremath{\Lambda\mathrm{CDM}}\xspace}
\newcommand{\logm}{\ensuremath{\log_{10}(m_{\rm a}[\si{\electronvolt}])}\xspace}
\newcommand{\oax}{\ensuremath{\Omega_\mathrm{a}}\xspace}
\newcommand{\om}{\ensuremath{\Omega_\mathrm{m}}\xspace}
\newcommand{\sigmaeight}{\ensuremath{\sigma_8}\xspace}
\newcommand{\Seight}{\ensuremath{S_8}\xspace}

\usepackage{natbib,twoopt}
\usepackage{xcolor}
\defcitealias{Ghirardini2024}{G24}
\makeatletter
\newcommandtwoopt{\citeads}[3][][]{\href{http://adsabs.harvard.edu/abs/#3}%
    {\def\hyper@linkstart##1##2{}%
     \let\hyper@linkend\@empty\citealp[#1][#2]{#3}}}
  \newcommandtwoopt{\citepads}[3][][]{\href{http://adsabs.harvard.edu/abs/#3}%
    {\def\hyper@linkstart##1##2{}%
     \let\hyper@linkend\@empty\citep[#1][#2]{#3}}}
  \newcommandtwoopt{\citetads}[3][][]{\href{http://adsabs.harvard.edu/abs/#3}%
    {\def\hyper@linkstart##1##2{}%
     \let\hyper@linkend\@empty\citet[#1][#2]{#3}}}
  \newcommandtwoopt{\citeyearads}[3][][]%
    {\href{http://adsabs.harvard.edu/abs/#3}
    {\def\hyper@linkstart##1##2{}%
     \let\hyper@linkend\@empty\citeyear[#1][#2]{#3}}}
\makeatother
\usepackage{graphicx}
\usepackage{multirow}
\usepackage{siunitx}
\usepackage{txfonts}

\DeclareSIUnit \parsec {pc}

\title{The SRG/eROSITA All-Sky Survey}
\subtitle{Constraints on ultralight axion dark matter through galaxy cluster number counts}

\author{S.~Zelmer\inst{1},
E.~Artis\inst{1},
E.~Bulbul\inst{1},
S.~Grandis\inst{2},
V.~Ghirardini\inst{3},
A.~von der Linden\inst{1,4}
Y.~E.~Bahar\inst{1}, 
F.~Balzer\inst{1},
M.~Br\"uggen\inst{5},
I.~Chiu\inst{6}, 
N.~Clerc\inst{7},
J.~Comparat\inst{1}, 
F.~Kleinebreil\inst{2},
M.~Kluge\inst{1},
S.~Krippendorf\inst{8}, 
A.~Liu\inst{1},
N.~Malavasi\inst{1},
A.~Merloni\inst{1},
H.~Miyatake\inst{9,10,11}, 
S.~Miyazaki\inst{12}, 
K.~Nandra\inst{1},
N.~Okabe\inst{13}, 
M.~E.~Ramos-Ceja\inst{1}, 
J.~S.~Sanders\inst{1}, 
T.~Schrabback\inst{2}, 
R.~Seppi\inst{14}, 
J.~Weller\inst{1,15}, and
X.~Zhang\inst{1}
} 

\institute{
Max Planck Institute for Extraterrestrial Physics, Giessenbachstrasse 1, 85748 Garching, Germany
\and
Universit\"at Innsbruck, Institut f\"ur Astro- und Teilchenphysik, Technikerstr. 25/8, 6020 Innsbruck, Austria
\and
INAF, Osservatorio di Astrofisica e Scienza dello Spazio, via Piero Gobetti 93/3, I-40129 Bologna, Italy
\and
Department of Physics and Astronomy, Stony Brook University, Stony Brook, NY 11794, USA
\and
Universität Hamburg, Hamburger Sternwarte, Gojenbergsweg 112, 21029 Hamburg, Germany
\and
Department of Physics, National Cheng Kung University, 70101 Tainan, Taiwan
\and
IRAP, Université de Toulouse, CNRS, UPS, CNES, F-31028 Toulouse, France
\and
University of Cambridge, Cavendish Laboratory and DAMTP, Cambridge CB3 0WA, United Kingdom
\and
Kobayashi-Maskawa Institute for the Origin of Particles and the Universe (KMI), Nagoya University, Nagoya, 464-8602, Japan
\and
Institute for Advanced Research, Nagoya University, Nagoya 464-8601, Japan
\and
Kavli Institute for the Physics and Mathematics of the Universe (WPI), The University of Tokyo Institutes for Advanced Study (UTIAS), The University of Tokyo, Chiba 277-8583, Japan
\and
Subaru Telescope, National Astronomical Observatory of Japan, 650 N Aohoku Place Hilo HI 96720 USA
\and
Department of Physical Science, Hiroshima University, 1-3-1 Kagamiyama, Higashi-Hiroshima,Hiroshima 739-8526, Japan
\and
Department of Astronomy, University of Geneva, Ch. d’Ecogia 16, CH-1290 Versoix, Switzerland
\and
Universit\"ats-Sternwarte, Faculty of Physics, LMU Munich, Scheinerstr. 1, 81679 M\"unchen, Germany
}
\date{\today}

\titlerunning{Constraints on ultralight axion dark matter through galaxy cluster number counts}
\authorrunning{Zelmer et al.}

\begin{document}

\abstract{Ultralight axions are hypothetical scalar particles that influence the evolution of large-scale structures of the Universe. Depending on their mass, they can potentially be part of the dark matter component of the Universe as candidates commonly referred to as fuzzy dark matter. While strong constraints have been established for pure fuzzy dark matter models, the more general scenario where ultralight axions constitute only a fraction of the dark matter has been limited to only a few observational probes. In this work, we use the galaxy cluster number counts obtained from the first All-Sky Survey (eRASS1) of the SRG/\erosita mission together with gravitational weak lensing data from the Dark Energy Survey, the Kilo-Degree Survey, and the Hyper Suprime-Cam to constrain the fraction of ultralight axions in the mass range $10^{-32}~\si{\electronvolt}$ to $10^{-24}\,\si{\electronvolt}$. We put upper bounds on the ultralight axion relic density \oax in independent logarithmic axion mass bins by performing a full cosmological parameter inference. We find an exclusion region in the intermediate ultralight axion mass regime with the tightest bounds reported so far in the mass bins around $m_\mathrm{a}=10^{-27}~\si{\electronvolt}$ with $\oax < 0.0035$ and $m_\mathrm{a}= 10^{-26}~\si{\electronvolt}$ with $\oax < 0.0079$; both are at a 95\% confidence level. When combined with cosmic microwave background probes, these bounds are tightened to $\oax< 0.0030$ in the $m_\mathrm{a}=10^{-27}~\si{\electronvolt}$ mass bin and $\oax < 0.0058$ in the $m_\mathrm{a}= 10^{-26}~\si{\electronvolt}$ mass bin, with both at a 95\% confidence level. This is the first time that constraints on ultralight axions have been obtained using the growth of structure measured by galaxy cluster number counts. These results pave the way for large surveys, which can be utilized to obtain tight constraints on the mass and relic density of ultralight axions with better theoretical modeling of the abundance of halos.}


\maketitle

\section{Introduction}
\label{sec:introduction}

According to the Lambda cold dark matter ($\Lambda$CDM) paradigm, the Universe is composed of two species of matter: baryonic matter and cold dark matter. The latter's potential interactions with the fundamental constituents of the standard model of particle physics are below the sensitivity of modern experiments, in particular its couplings with photons. Additionally, the $\mathrm{\ LambdaCDM}$ includes a dark energy component responsible for the accelerated expansion of the Universe, which is parameterized by the cosmological constant $\Lambda$. Various cosmological probes have provided tight constraints on the dark matter, baryonic matter, and dark energy content of the Universe, for example, through the analysis of the cosmic microwave background (CMB) \citep{Planck2015, Planck2020, WMAP2013}, Type Ia supernovae \citep{Abbott2019, DESSNIa2024, Scolnic2022, Brout2022}, cluster abundance (\citealp[][\citetalias{Ghirardini2024} hereafter]{Ghirardini2024}, \citealp{Bocquet2024}, \citealp{Costanzi2021}, \citealp{Lesci2022}), weak lensing shear \citep{Asgari2021, Abbott2022, Amon2022, Miyatake2023, Dalal2023}, and galaxy clustering \citep{Zhao2022, DESI2024}.

Despite the constraints on the density of dark energy and dark matter obtained by state-of-the-art cosmological studies, their nature remains elusive.
Consequently, potential extensions of $\Lambda$CDM with viable dark matter candidates have been explored using observations of the large-scale structure, such as axion-like particles (ALPs), and self-interacting dark matter \citep{Randall2008, Conlon2017}. The ALPs are promising potential dark matter candidates. 
Depending on their properties, ALPs can also partially explain dark energy. ALPs are scalar fields with typically tiny masses, compared to known elementary particles, that move in a periodic potential. Historically, the so-called quantum chromodynamics (QCD) axion was proposed by \cite{Peccei1977a, Peccei1977b} to solve a fine-tuning problem of QCD, known as the strong charge parity (CP) problem. Although predicted by QCD, the strong CP problem refers to unobserved violations of simultaneous charge conjugation (C) and parity (P). Particles with the same underlying physics naturally occur in string theory due to the compactification of large extra dimensions and are called axions or ALPs \citep{Svrcek2006}. The typical particle masses can span the range from $10^{-33}~\si{\electronvolt}$ to $10^{-5}~\si{\electronvolt}$. In principle, multiple ALPs with a large spectrum of mass might exist in string theory, called the axiverse \citep{Arvanitaki2010}. Upper bounds on the number of ALP species have been found by superradiance effects of stellar and supermassive black holes \citep{Stott2018} and supernovae \citep{Gendler2024}. ALPs with masses $m_\mathrm{a}\gtrsim 10^{-22}~\si{\electronvolt}$ are candidates to compose a large fraction of dark matter, while ALPs with masses $m_\mathrm{a}\lesssim 10^{-33}~\si{\electronvolt}$ represent candidates for dark energy. 

In the dark energy scenario ($m_\mathrm{a}\sim 10^{-33}~\si{\electronvolt}$), the axion field exhibits slow-roll behavior due to the small mass of ALPs. As it is still rolling today, the axion field has not yet started to oscillate around the minimum of its potential. Thus, it behaves similarly to a fluid with negative pressure, which is similar to dark energy \citep{Hlozek2015, Hlozek2018, Passaglia2022}. For masses $m_\mathrm{a} \lesssim 10^{-33}~\si{\electronvolt}$, the axion field is frozen in, and only its potential energy contributes to the vacuum energy of the Universe. This scenario is indistinguishable from the \lcdm dark energy with the equation of state $w=-1$.

In the case of small masses around $10^{-24} \,\si{\electronvolt}$, the bosonic nature and negligible self-couplings of ALPs enable them to form Bose-Einstein condensates (BECs) on scales determined by their thermal de Broglie wavelength. The simulations by \cite{schive14a} have demonstrated the formation of core-like structures in the center of halos, whose size could be used to obtain constraints on the ALP mass from dwarf galaxies. Over the past decades, a number of simulations of ALPs as dark matter candidates (also called wave-like dark matter) have been performed with increasing resolution and while assuming a cosmology with a dark matter sector fully comprised by ALPs \citep[e.g.,][]{Mocz2018, Nori2018, May2021, May2023}. With an increasing precision of ALP mass constraints, mixed ALP-cold dark matter simulations became relevant since they can be used to constrain the fraction of ALPs in the Universe \citep{Schwabe20, Lague2024, Dome2025}. In this axion mass regime, the thermal de Broglie wavelength extends to the cosmological scales, i.e., several kilo- or megaparsecons, making the axion condensates observable in studies of the matter power spectrum. A BEC on these scales forms large core-like structures, smooths out the dark matter distribution, and suppresses the formation of small-scale halos. ALPs have been proposed as a potential solution to the small-scale crisis in cosmology, particularly for addressing discrepancies in structure formation at galactic and subgalactic scales. However, the existence of the small-scale crisis itself has been questioned, with alternative explanations suggesting that baryonic physics or observational limitations may account for the perceived discrepancies \citep[][]{Nori2018, Bullock2017, DeLaurentis2022}. We refer to these cosmologically interesting low-mass axions with masses $m_\mathrm{a} \lesssim 10^{-24}~\si{\electronvolt}$ as ultralight axions (ULAs). In the literature, the resulting dark matter is referred to as fuzzy dark matter due to its smoothing effect \citep[e.g.,][]{Hu2000, Hui2017}.

Constraining the coupling constant between ALPs and photons is also feasible, but it requires direct X-ray observations of the intracluster medium (ICM), which have the potential to detect a possible interaction between axions and photons. Such an interaction is only possible in magnetic fields since the axion coupling term is proportional to the scalar product of electric and magnetic fields. Measurements of spectral features in galaxy clusters have been used to constrain the properties of dark matter candidates such as ALPs and decaying dark matter \citep{Bulbul2014, Conlon2017, Reynolds2020}. Unlike the studies performed with the probes of structure growth, these analyses can determine bounds on the axion-photon coupling constant and mixing angles, but they cannot constrain the relic density of dark matter candidates.

The viable mass ranges for ULAs and abundances constituting dark matter and dark energy can be constrained through observations of the large-scale structure as well as the CMB and subgalactic scales. In the dark matter regime, where the ULAs make up all of the dark matter by assumption, strong bounds on the ULA parameter space have been reported in the literature. Lyman-alpha forest observations place a lower-mass bound of $2 \times 10^{-20} \,\si{\electronvolt}$, at a 95\% confidence level \citep{Rogers2021}. The observations of ultrafaint dwarf galaxies tighten this bound to $3 \times 10^{-19} \,\si{\electronvolt}$, at a 99\% confidence level \citep{Dalal2022}, while superradiance by supermassive black holes can be used to exclude a window of masses around $7\times 10^{-20} \,\si{\electronvolt}$ \citep{Arvanitaki2015, Cardoso2018, Stott2018, Stott2020, Hoof2024}.

The possibility that ULAs may comprise only a fraction of the total abundance of dark matter nevertheless remains. For the first time, CMB data from the Planck satellite \citep{Planck2016b} provided strong bounds in the mass regime $10^{-32} \,\si{\electronvolt}$ to $10^{-25} \,\si{\electronvolt}$ by performing a binned analysis in different ULA mass bins. They found varying upper bounds on the ULA abundance, reaching down to $\oax \lesssim 0.01$ in the intermediate mass range of $10^{-30}-10^{-27}$~eV \citep{Hlozek2015, Hlozek2018}. The same mass regime has also been constrained by \cite{Lague2022} and \cite{Rogers2023} using \planck 2018 \citep{Planck2020} and Baryon Oscillation Spectroscopic Survey (BOSS; \citep{Alam2017}) data, improving existing constraints. \cite{Kobayashi2017} used the Lyman-$\alpha$ forest to place bounds on the fraction of ULAs in the higher-mass regime of $10^{-23} \,\si{\electronvolt}$ to $10^{-20} \,\si{\electronvolt}$, and \cite{Winch2024} closed the gap by using UV-bright galaxy abundance to put an upper bound on the ULA fraction in the mass regime $10^{-26} \,\si{\electronvolt}$ to $10^{-23} \,\si{\electronvolt}$. They find that ULAs cannot make up more than 22\% of all dark matter in this mass range.

In this work, we use for the first time the observed abundance of galaxy clusters to constrain the parameter space of ULAs. Since the formation of halos is influenced by bosonic condensation, we expect cluster counts to be sensitive to ULAs \citep{Diehl2021}. We particularly focus on the mass regime $10^{-32} \,\si{\electronvolt}$ to $10^{-24} \,\si{\electronvolt}$. Clusters of galaxies trace the highest peaks in the low-redshift matter density field through their number counts and therefore serve as an important late-time probe of the structure formation in the Universe. Measurements of their number count as a function of mass and redshift bin per unit volume, the so-called cluster halo mass function (HMF), is a sensitive probe of cosmology. The cluster mass function can constrain the ULA mass and abundance parameter space since the growth of structure and bosonic condensation influence the formation of halos.

To this end, we used the clusters of galaxies detected by the extended ROentgen Survey with an Imaging Telescope Array (\erosita) on board the Spektr-RG (SRG) mission \citep{Sunyaev2021} during its first All-Sky Survey in the western Galactic hemisphere (eRASS1). \erosita, launched in 2019, scans the sky in the soft X-ray band with its highest sensitivity in the 0.2 to 2.3 keV energy band \citep{Predehl2021}. eRASS1 provides different cluster catalogs depending on purity and completeness requirements. The primary eRASS1 catalog comprises 12,247 galaxy clusters, with a sample purity of 86\% and a coverage of 12,791 $\mathrm{deg}^2$ \citep{Bulbul2024, Kluge2024}. In this work, we use the cosmology sample containing 5,259 securely detected clusters in the redshift range of $0.1 \leq z \leq 0.8$, which have a sample purity of 96\%. The eRASS1 cosmological sample has provided competitive results for standard cosmologies (see \citetalias{Ghirardini2024} for the details of the analysis) as well as modified gravity \citep{Artis2024} and the growth of structures \citep{Artis2025}. The eRASS1 cosmology sample is the largest ICM-selected galaxy cluster sample ever used to infer cosmology from number counts. The covered cluster mass range extends down into the galaxy group regime, opening the possibility of probing cosmological models sensitive to below-cluster scales. The presence of ULAs has the strongest imprint on the number density of low-mass halos, making the eRASS1 cosmology sample a promising tool to constrain the properties of ULAs. In contrast to the main cosmological inference \citetalias{Ghirardini2024}, in this work, to constrain the ULA parameter space, we used the Boltzmann solver {\texttt {axionCAMB}} instead of {\texttt{CAMB}} to compute a matter power spectrum that is sensitive to two ULA parameters, particularly the mass of the ULA, $m_\mathrm{a}$, and the energy density fraction of ULAs, $\oax$, in units of the critical density of the Universe. 

This paper is organized as follows. A more detailed overview of the ULAs and their effect on cosmology is given in Sect.~\ref{sec:axion_cosmology}, including their influence on structure formation. The underlying cosmological sample of the eRASS1 survey and the optical and weak lensing data used for the analysis are summarized in Sect.~\ref{sec:multi-wavelength_data}. Section~\ref{sec:methodology} reviews the eRASS1 cluster cosmological pipeline for Bayesian inference in light of the ULA parameters. The results are presented and discussed in Sects.~\ref{sec:results} and \ref{sec:conclusion}. For better readability, we use natural units $c\equiv \hbar \equiv 1$ throughout this work.

\section{Ultralight axion cosmology}
\label{sec:axion_cosmology}

In this section, we first provide a review of the motivation for ULAs as dark matter candidates. We then explain the implications of their existence on the hierarchical structure formation and cosmology. 

\subsection{Physical origin of axions}
\label{sec:physical_origin_of_axions}

Axions are hypothetical pseudoscalar particles with typically low mass (between $10^{-33} \,\si{\electronvolt}$ and $10^{-5} \,\si{\electronvolt}$) and weak couplings to the Standard Model particles (i.e., quadratic coupling terms can be neglected). Pseudoscalar hereby refers to the property that the particle behaves like a scalar under transformations, except parity transformations, under which it switches signs. Although the axion was originally introduced in QCD to solve the strong CP problem explicitly, the term has been extended to a number of (pseudo) Nambu-Goldstone bosons. Nambu-Goldstone bosons are particles that occur as massless degrees of freedom after breaking a spontaneous symmetry. Axions are pseudo-Nambu-Goldstone bosons as they acquire a very small mass via further symmetry breaking. These particles appear in various theories of high energy physics such as string theory \citep[e.g.,][]{Svrcek2006, Conlon2006, Asimina2010} or supersymmetric extensions of the Standard Model \citep[e.g.,][for review]{Jihn2010, Marsh2016}. To distinguish such general axions from the QCD axion, they are often referred to as ALPs or ultralight axions in the cosmologically relevant low-mass regime. The common properties of these particles are well described in section 2.2 in \cite{Marsh2016}. 

We heuristically illustrate axion physics using the QCD axion introduced by \cite{Peccei1977a, Peccei1977b} as a generic example. We refer the reader to \cite{Marsh2016} for further details. A non-Abelian $\mathrm{SU}(3)$ gauge theory with a gluon gauge field $G_{\mu \nu}^{\rm a}$ naturally exhibits a topological term with some coupling $\theta$ in its Lagrangian density, which is of the form

\begin{equation}
\mathcal L \supset \frac{\theta}{32\pi^2}G_{\mu \nu}^a \tilde{G}^{\rm a \mu \nu},
\end{equation}

\noindent where $\tilde{G}^{\rm a \mu \nu} = \epsilon^{\mu \nu \alpha \beta} G^{\rm a}_{\alpha \beta} / 2$ is the dual of the gluon gauge field tensor,  $\mu$ and $\nu$ refer to spacetime coordinates and $a$ is some color index. This term is CP-violating and cannot be transformed away. If QCD is CP-violating, it leads to a non-vanishing neutron electric dipole moment $d_{\rm n} \approx \theta \, (3.6 \times 10^{-16} e \, \rm{cm})$ \citep{Crewther1979}. However, experiments place strong constraints on the neutron electric dipole moment \citep{Abel2020}, which results in a fine-tuned value of $\theta \lesssim 10^{-10}$. In other words, QCD appears to be non-CP-violating, although it should. This is known as the strong CP problem.

\cite{Peccei1977a, Peccei1977b} solved this problem by generalizing the coupling constant $\theta$ to a scalar field that naturally cancels any contribution to the CP-violating term. To this end, they introduced a new global $\mathrm{U}_{\rm{PQ}}(1)$ symmetry, known as the Peccei-Quinn symmetry. The spontaneous breaking of this symmetry at a high energy scale $f_{\rm{a}}$ leads to a vacuum manifold that is isomorphic to the one-sphere $S^1$. The emerging Nambu-Goldstone boson (which will become the axion later) is an angular degree of freedom in an effectively flat potential. Thus, at this point, the new particle is massless. At some lower energy scale, $\Lambda_{\rm{a}}$ (not to be confused with the cosmological constant $\Lambda$), non-perturbative QCD effects (e.g., instantons, topological effects) break the flatness of the vacuum manifold, the massless Nambu-Goldstone boson lives in. Due to the angular nature of the Nambu-Goldstone boson, only a discrete angular shift symmetry is preserved. Ultimately, the Nambu-Goldstone boson acquires a mass and is canonically called an axion. Simplifying further, the field rolls into the circular minimum of a sombrero-like potential during the first spontaneous symmetry breaking. The Nambu-Goldstone boson is the angular degree of freedom that moves freely along the minimum of the potential. The angular shift symmetry of this potential is then broken by nonperturbative effects, leading to local minima and maxima along the former circular minimum of the potential. Particles moving in a potential with minima acquire a mass; thus, the Nambu-Goldstone boson becomes a massive axion particle. 

The axion mass is related to the symmetry-breaking scales $f_{\rm a}$ and $\Lambda_{\rm a}$. A typical but not unique effective axion potential can take the form

\begin{equation}
V(\phi) = \Lambda_{\rm a}^4 \left[ 1-\cos \left( \frac{\phi}{f_{\rm a}} \right) \right].
\end{equation}

Considering small oscillations around one of the local minima gives the effective mass of the axion by approximating the local environment of the minimum with a quadratic function:

\begin{align}
V(\phi\approx \phi_{\rm {min}}) & \approx \frac{1}{2}\left(\frac{\Lambda_{\rm a}^2}{f_{\rm a}}\right)^2\phi^2 + \Lambda_{\rm a}^4\mathcal{O}\left(\frac{\phi^4}{f_{\rm a}^4}\right) \\ & = \frac{1}{2}m_{\rm a}^2\phi^2 + \left({\text{self-interactions}}\right).
\end{align} 

The mass of the axion can be read off as $m_{\rm a} = \Lambda_{\rm a}^2/f_{\rm a}$. Typical values of $f_{\rm a}$ lie between $10^{9}\,\si{\giga\electronvolt}$ and $10^{17}\,\si{\giga\electronvolt}$ in the QCD case and, in the general case, above $10^{10}\,\si{\giga\electronvolt}$ with typical values around the grand unified theory (GUT) scale at $10^{16} \,\si{\giga\electronvolt}$ \citep{Marsh2016}. In general, self-interactions are thus suppressed by $m_\mathrm{a}^2/f_\mathrm{a}^2 \lesssim 10^{-90}$ for the ULA mass regime considered in this work, and we make the assumption that self-interactions are negligible in our analysis.

The underlying mechanisms of the QCD axion also apply to axions in string theory models. However, in string theory, the periodicity of the axion potential originates from the compactification of extra dimensions, i.e., extra dimensions are wrapped up, which makes them unobservable by macroscopic observers. The need for extra dimensions in string theory, which have not been observed yet, leads to a natural production mechanism of axion particles with a potentially wide range of possible axion masses.

\begin{figure*}[h!]
\includegraphics[width=0.495\textwidth, clip]{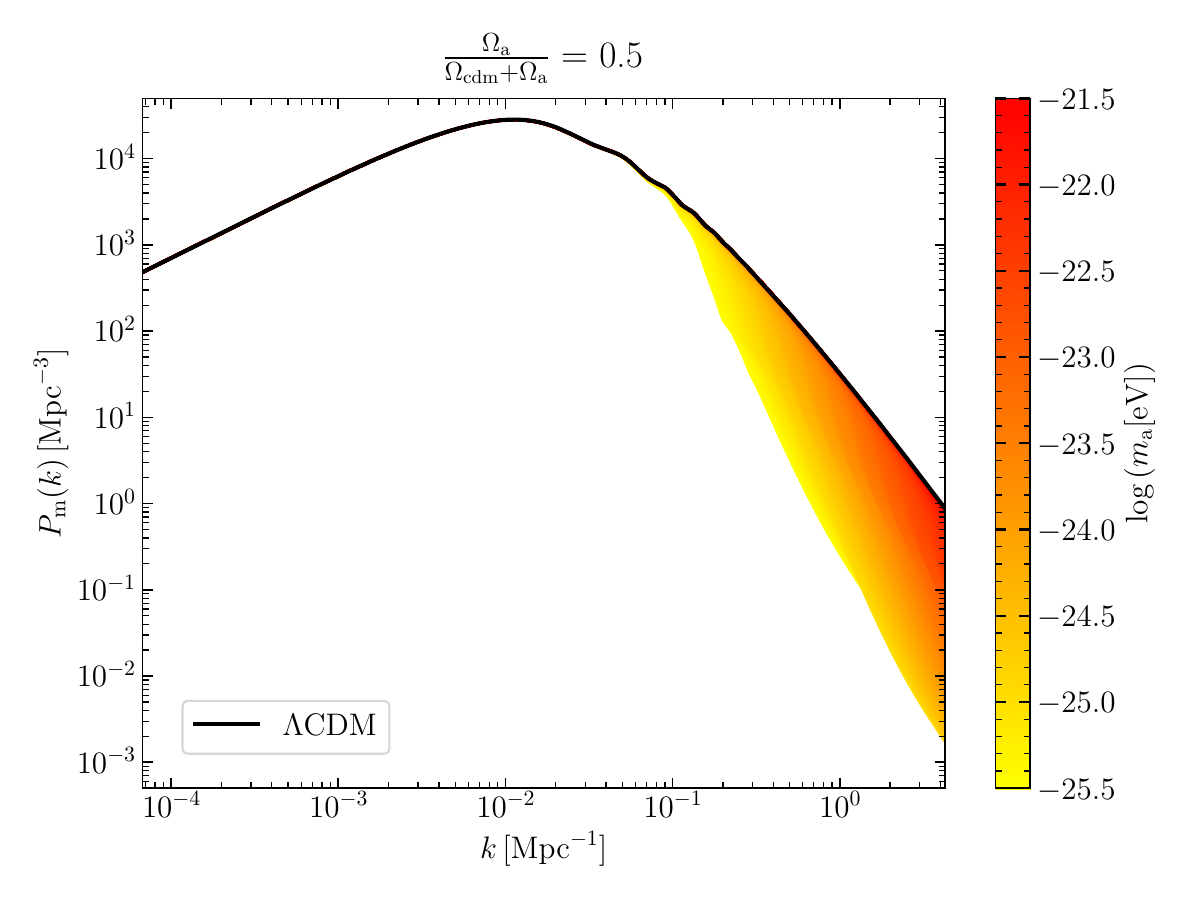}
\includegraphics[width=0.495\textwidth]{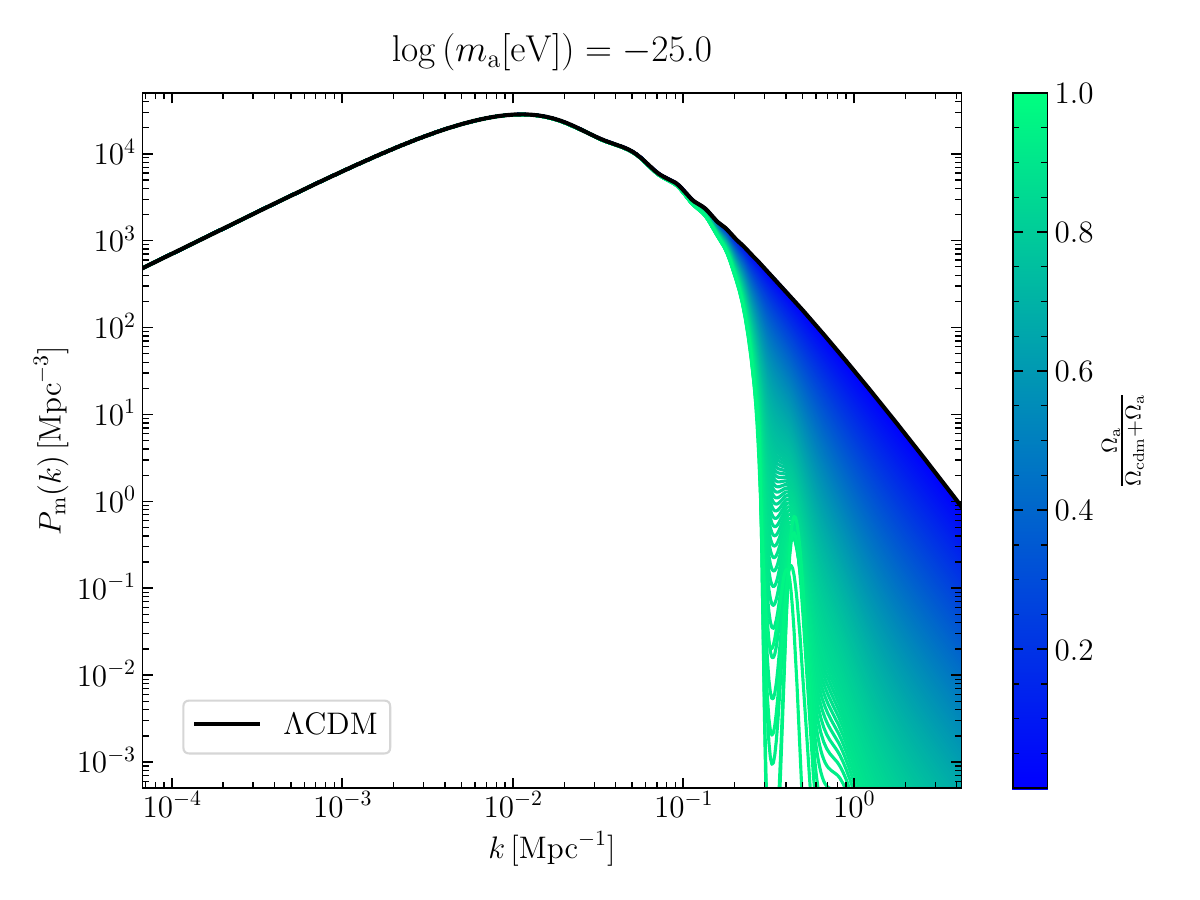} 
\caption{\label{fig:matterpowerspectrum}Effect of ULAs on the matter power spectrum recovered assuming the \cite{Planck2020} cosmological values ($\Omega_\Lambda \approx 0.7$, $\Omega_\mathrm{m}+\oax \approx 0.3$) at $z=0.1$. On the left, the change in the matter power spectrum with varying ULA mass is shown when the relative ULA density is fixed to $\oax/(\Omega_\mathrm{cdm}+\oax)=0.5$. The panel on the right displays the dependence of the power spectrum on varying ULA abundance at a fixed ULA mass of $m_{\rm a}=10^{-25} \,\si{\electronvolt}$. The concordance \lcdm\ model is shown in a black curve on both panels. Small-scale suppression becomes stronger with decreasing ULA mass and increasing ULA abundance. For higher ULA masses of $m_{\rm a} \gg 10^{-22} \,\si{\electronvolt}$, the matter power spectrum becomes indistinguishable from the one in a $\Lambda$CDM cosmology, as is the case for $\oax \rightarrow 0$.}
\end{figure*}

\subsection{Cosmological implications of ultralight axions}
\label{sec:cosmological_implications_of_axions}

This section reviews the implications of an ultralight axion fluid on cosmology. For in-depth discussion and derivations, we refer the reader to the derivations presented in \cite{Marsh2016}, \cite{Hlozek2015}, and \cite{Hlozek2018}. The ULA field enters the Einstein-Hilbert action of general relativity as a massive scalar field in curved spacetime. On a Friedmann-Lema\^itre-Robertson-Walker metric background, the Klein-Gordon equation (equation of motion) of the ULA is

\begin{equation}
\label{eq:eom}
\ddot \phi + 3H \dot \phi + m_{\rm a}^2\phi^2 = 0.
\end{equation}

Here, $H(t) = \dot a(t)/a(t)$ is the Hubble parameter, $\phi$ is the ULA field, and dots refer to derivatives in time. This equation describes a harmonic oscillator in $\phi$ with a damping term proportional to the Hubble parameter $H$. For this illustration, we neglect any source terms from other cosmological fluid components or self-interactions, which would enter the equation on the right-hand side and replace the zero. For $H \ll m_{\rm a}$, i.e., if the ULA mass is orders of magnitude larger than the Hubble parameter, the damping term is negligible, and the ULA field oscillates freely. If the ULA's mass is sufficiently large, this describes the matter-like late-time behavior of the ULA. For the opposite case, $H \gg m_{\rm a}$, the field is overdamped and frozen in, showing the same behavior as dark energy. The transition between the two regimes defines a scale $a_{\rm{osc}}$ which can be approximated by $H(a_{\rm{osc}}) \approx m_{\rm a}$. In this regime, the ULA is slow-rolling. Thus, a single ULA species may transition from a dark energy-like behavior to a dark matter-like behavior. As the universe evolves in the opposite direction from matter-dominated to dark energy-dominated era, no single ULA species can describe both dark matter and dark energy. We therefore distinguish two distinct phenomenological regimes for ULAs: the dark matter regime, where they behave as dark matter, and the dark energy regime, where they show the same observed effect of dark energy. We review the two regimes in the following subsections.

\subsection{Dark matter regime of ultralight axions}

Introducing perturbations to the ULA field and choosing a gauge yields an equation for the evolution of perturbations in terms of the ULA overdensity $\delta_{\rm a} = \delta \rho_{\rm a} / \rho_{\rm a}$. In the simple case of a ULA-dominated universe, it takes the form

\begin{equation}
\label{eq:axionoverdensities}
\ddot \delta_{\rm a} + 2H\dot \delta_{\rm a} + \left( \frac{k^2c_{\rm {s, a}}^2}{a^2} - 4\pi G \rho_{\rm a} \right)\delta_{\rm a} = 0,
\end{equation}

\noindent  where $k$ is the wave number, $a$ the scale factor, $\rho_{\rm a}$ describes the background density field of the ULA, $\delta \rho_{\rm a}$ its perturbation. Other potential fluid components would add source terms to the equation. The effective sound speed of ULAs ($c_{\rm {s,a}}$) becomes

\begin{equation}
\label{eq:axionsoundspeed}
c_{\rm {s, a}}^2 = \frac{k^2}{k^2 + 4m_{\rm a}^2a^2}.
\end{equation}

Equation~\ref{eq:axionoverdensities} defines a scale-dependent Jeans scale $k_{\rm J}$ for ULAs. As described in \cite{Marsh2016}, large-scale perturbations ($k<k_{\rm J}$) grow and behave similarly to CDM. Small-scale perturbations ($k>k_{\rm J}$) oscillate, making them distinct from CDM perturbations on the same scales. As described in \cite{Marsh2012}, modes entering the horizon while the ULA sound speed in Eq.~\ref{eq:axionsoundspeed} is near the speed of light will suppress structure formation, while modes entering the horizon while the ULA sound speed is approaching zero will cluster like ordinary cold dark matter. Effectively, this translates to suppressing the matter power spectrum on small physical scales (large values of the wave number $k$).

\begin{figure}[h!]
    \centering
    \includegraphics[width=0.5\textwidth]{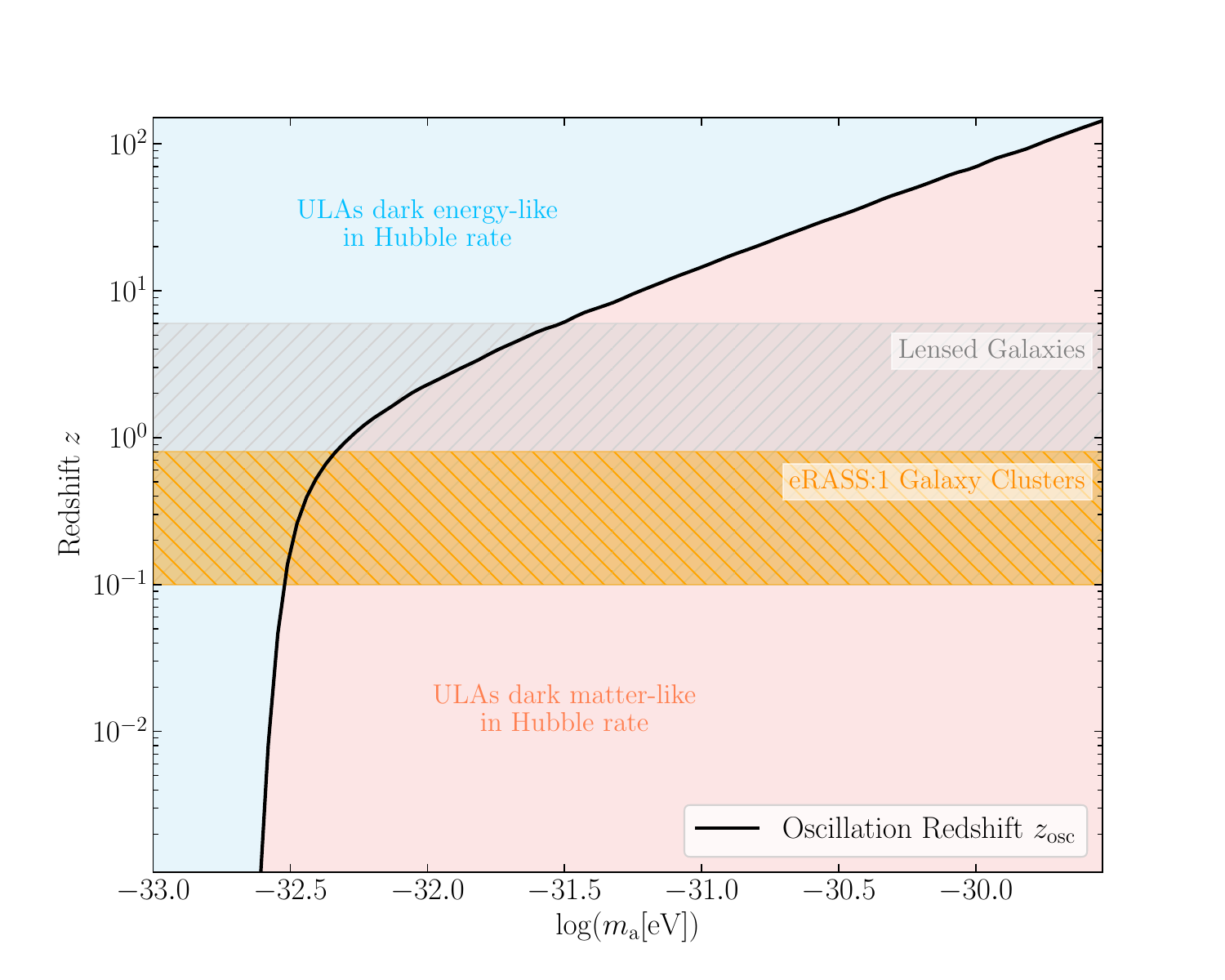}
    \caption{Demonstration of the ULA mass and redshift dependent regime change in the Hubble rate. In the pink regime, ULAs contribute to the Hubble rate e in a manner similar to dark matter; in the blue region, ULAs contribute to the Hubble rate in a dark energy-like fashion. The orange shaded region shows the eRASS1 cluster redshift range, while the gray shaded region shows the redshift range of lensed galaxies from the DES, HSC, and KiDS data, described in Sect.~\ref{sec:wl_data}. Above a ULA mass of $m_\mathrm{a} \sim 10^{-31.5}~\si{\electronvolt}$, ULAs should be modeled as dark matter in the Hubble rate, resulting in correct distance measurements during the parameter inference. For ULAs with masses below $m_\mathrm{a} \sim 10^{-31.5}~\si{\electronvolt}$, the distance measurements are affected by the regime change via the Hubble rate, where ULAs are treated as dark energy.}
    \label{fig:regimechange}
\end{figure}

The properties of ULAs and their cosmological effects can be described with two quantities: The particle mass of the ULA $m_{\rm a}$, and the relative energy density of ULAs with respect to the critical density of the universe, $\Omega_{\rm {ax}} = \rho_{\rm {ax}}/\rho_{\rm {crit}}$, also known as the relic density of ULAs. Here, the critical density of the universe is defined as  $\rho_\mathrm{crit} = 3H_0^2/(8\pi G)$. The \lcdm cosmology is recovered when $\oax \rightarrow 0$. Furthermore, the ULAs' suppression imprint on the matter power spectrum becomes negligible for high ULA masses $m_\mathrm{a} \gg 10^{-22}~\si{\electronvolt}$, effectively recovering a \lcdm cosmology in the context of this analysis. Fig.~\ref{fig:matterpowerspectrum} demonstrates the suppression effect of ULA dark matter with different logarithmic particle masses \logm and relic densities \oax on the matter power spectrum. Suppression of the power can be seen on small scales (large $k$), whereby the effect is stronger with higher ULA fractions or lower ULA masses. 

In this work, we explore whether a fraction of dark matter is composed of ULAs of an unknown mass in the dark matter regime. We still need to postulate a cosmological constant, dark energy, in this context. As we assume a flat universe, the dark energy density is given by $\Omega_\Lambda = 1 - \Omega_\mathrm{m} -  \oax$, while we fit for the total matter density $\Omega_\mathrm{m}$ and the ULA density \oax. The results are provided in Sect.~\ref{sec:results}.

\subsection{Dark energy regime of ultralight axions}
\label{sec:dark_energy_regime}

If the ULA is so light that its mass becomes comparable with or smaller than the Hubble parameter $H_0$, the oscillation scale $a_\mathrm{osc}$ approaches or even exceeds the current scale factor of the universe $a(z=0)=1$, indicating that the ULA field has not started oscillating yet. Indeed, this kind of field behaves as dark-energy-like, which can be heuristically illustrated with the equation of motion. Neglecting the $m_\mathrm{a}^2\phi^2$ term in Eq.~\ref{eq:eom} and assuming that $H(t)$ varies slowly enough in time exhibits exponential modes for the ULA field $\phi$. This behavior is associated with an equation of state $w_\mathrm{ax} \approx -1$ via the density and pressure of the ULA field. In terms of the de Broglie wavelength associated to the ULA BEC, in the extreme case of $H_0 \approx m_\mathrm{a}$ which occurs around ULA masses of $m_\mathrm{a} \approx 10^{-33}\,\si{\electronvolt}$ the de Broglie wavelength is comparable to the size of the universe leading to an overall energy shift, typically associated to a dark energy-like component. However, even at higher ULA masses, as long as the de Broglie wavelength of the ULA BEC exceeds its Jeans scale, ULAs cannot form halos and are thus still referred to as dark energy-like ULAs.
Furthermore, if the ULA BEC de Broglie wavelength is comparable to the size of the universe, this affects the Hubble rate and therefore also the distance measurements. 
This results in a redshift-dependent transition in the Hubble rate, with the nature of the transition dependent on the ULA mass. Wherever the de Broglie wavelength exceeds the size of the universe, ULAs should be modeled as a dark energy component in the Hubble rate. In the opposite case of a de Broglie wavelength smaller than the size of the universe, ULAs should be modeled as a dark matter component in the Hubble rate. The transition between the two regimes is determined by the oscillation redshift.
Fig.~\ref{fig:regimechange} illustrates the dependence of the oscillation redshift on the ULA mass. We emphasize here that the regime change in the Hubble rate is not related to the typically referenced regime change between dark energy-like ULAs' and dark matter-like ULAs' contribution to halo formation, which is determined by the respective Jeans scale.

\section{Multiwavelength data}
\label{sec:multi-wavelength_data}

\subsection{eROSITA cluster catalog and X-ray observable}

This work uses the cosmology sample compiled from the first eROSITA All-Sky Survey \citep{Merloni2024, Bulbul2024, Kluge2024}. This sample is constructed by employing an extent likelihood cut of $\mathcal{L}_\mathrm{ext}>6$ and a redshift range of $0.1 \leq z \leq 0.8$. The publicly available DESI Legacy Survey Data Release 10 (LS DR10) data are used for the confirmation, identification, and photometric redshift measurements of the X-ray-selected cluster candidates detected in the eROSITA All-Sky Survey's western Galactic hemisphere \citep{Kluge2024}. The common footprint between eRASS1 and the LS DR10-South footprint is 12,791~deg$^2$, and contains 5,259 securely confirmed clusters \citep{Kluge2024}. Its purity level is estimated to be 96\%.

The eROSITA X-ray data are reprocessed with the \esass software as described in \citet{Merloni2024} by applying a correction for the Galactic absorption and a more accurate ICM and background modeling. The details of the X-ray processing are provided in \citet{Bulbul2024, Liu2022}. The X-ray observable count rate in the observer frame, which demonstrates a tight correlation with weak lensing shear data and is related to selection, is used as a mass proxy for this work, similar to the method adopted in \citetalias{Ghirardini2024}.

\subsection{Weak lensing data}
\label{sec:wl_data}

In order to minimize the bias on galaxy cluster masses, we use weak gravitational lensing data to obtain reliable mass measurements for the calibration of the mass-observable scaling relations.
We used weak lensing measurements in the form of tangential reduced shear profiles extracted from the shapes of background galaxies in the Dark Energy Survey year three (DES~Y3) data \citep{Sevilla2021, Gatti2021} and the Kilo Degree Survey (KiDS), and the Hyper Suprime-Cam (HSC) Strategic Survey program \citep{Hildebrandt2021, Li2022}. The details of shape measurements around the detected eROSITA clusters and the weak lensing mass calibration details are provided in \cite{Grandis2024}, \cite{Kleinebreil2024} and \cite{Chiu2025}. Besides the reduced tangential shear profile, we also use estimates of the redshift distribution for the employed background sources, estimates of the measurement uncertainty on the reduced shear profile, and calibrations on the contamination of the background sample by cluster member galaxies.

The Dark Energy Survey shares a common sky area of 4,060 $\mathrm{deg}^2$ with \erosita, and its observations have been conducted in the $r$, $i$, and $z$ bands \citep{Sevilla2021}. The details of the application of the DES~Y3 shear maps to the eRASS1 galaxy clusters are presented in \cite{Grandis2024}. Tangential shear profiles were derived from the DES~Y3 shape catalog \citep{Gatti2021}. The analysis yielded 2,201 tangential shear profiles for eRASS1 galaxy clusters with a signal-to-noise ratio of 65.

The Hyper Suprime-Cam Subaru Strategic Program \citep{Aihara2018} is a wide and deep optical survey in the $g$, $r$, $i$, $z$, and $y$ bands. The three-year shape catalog (HSC~Y3) \citep{Li2022} was used to analyze galaxy cluster shear profiles in an overlapping sky area of $\approx 500~\mathrm{deg}^2$ between HSC~Y3 and eRASS1. This analysis yielded tangential shear profiles, lensing covariance matrices, and redshift distributions for 96 eRASS1 galaxy clusters. The total signal-to-noise is 40 \citep[for details see][]{Chiu2025, Okabe2025, Chiu2022}.

The Kilo-Degree Survey is an optical wide-field survey in the $u$, $g$, $r$, and $i$ bands, dedicated to delivering weak gravitational lensing as well as photometric redshift measurements \citep{Jelte2013}. We use the shear maps and photometric redshifts of the fourth data release of the Kilo-Degree Survey (KiDS-1000) over a sky area of $\approx 1,000~\mathrm{deg}^2$ \citep{Kuijken2019, Giblin2021, Hildebrandt2021, Angus2020}. The joint sky coverage with \erosita yielded reduced tangential shear maps for 237 eRASS1 galaxy clusters (101 in the KiDS-North field, 136 in the KiDS-South field) with a total signal-to-noise of 19 \citep[for details see][]{Kleinebreil2024}.

We highlight the importance of the overlap between eRASS1 clusters in KiDS and HSC (125 clusters) and in KiDS and DES (25 clusters). There is no overlap in the sky coverage of HSC and DES, so the reduced tangential shear maps deduced from KiDS-1000 provide an internal consistency check \citep[][\citetalias{Ghirardini2024}]{Kleinebreil2024}.

\section{Methodology}
\label{sec:methodology}

This section briefly presents the inference pipeline developed in \citetalias{Ghirardini2024} and explains the applied adaptations to consider the cosmological effects of introducing ULAs. To this end, we reviewed the observables, the HMF, the mass-observable scaling relations, and the weak gravitational lensing calibration.

The forward modeling cosmological pipeline is described in \citetalias{Ghirardini2024}. It takes into account selection effects through the selection function presented in \cite{Clerc2024} to model the completeness and mass-observable scaling laws calibrated with weak gravitational lensing data by DES~Y3, KiDS, and HSC~Y3, presented in \cite{Grandis2024, Kleinebreil2024, Chiu2025}, as well as a mixture model for eliminating care of possible contaminants \citep{Kluge2024}. The Bayesian analysis assumes a Poissonian likelihood for the galaxy cluster number counts. The pipeline is used for constraining a large set of cosmological parameters as well as scaling parameters simultaneously. 

\subsection{Statistical inference}
\label{sec:stat_inf}

This section summarizes the framework presented in \citetalias{Ghirardini2024} for the computation of the cluster count likelihood.
The cluster abundance, $n$, per units of true mass, $M$; true redshift, $z$; and solid angle (true sky positions are noted with $\mathcal{H}$) follows

\begin{equation}
    \label{eq:halo_mass_function}
    \frac{\mathrm{d}\,n}{\mathrm{d}\ln M\mathrm{d}\,z\mathrm{d}\,\mathcal{H}} = \frac{\rho_\mathrm{m,0}}{M} \frac{\mathrm{d}\ln \sigma^{-1}}{\mathrm{d}\ln M}f(\sigma)\frac{\mathrm{d}\,V}{\mathrm{d}\,z\mathrm{d}\,\mathcal{H}}, 
\end{equation}

\noindent where $\rho_\mathrm{m,0}$ is the matter density at present, $\sigma(R,z)$ is the root mean square density fluctuation defined below, $\mathrm{d}\,V/\mathrm{d}\,z\mathrm{d}\,\mathcal{H}$ is the differential comoving volume per redshift per steradian. Further, $f(\sigma)$ is the multiplicity function introduced by \citet{Tinker2008}, with its parameters assumed to be fixed and universal.
The root mean square density fluctuation, $\sigma$, is defined as the variance of a convolution of the square of a window function, $W$, with the matter power spectrum, $P$:

\begin{equation}
\label{eq:rms_density}
\sigma^2(z, R) = \int_{0}^{\infty} {\rm d}k \, k^2 P(k, z) |W(kR)|^2,
\end{equation}

where $k$ is the wave number of the considered mode and $R$ is the defining radius of the corresponding overdensity. For $W$, the standard top-hat window function is used, which defined as the Fourier transform of a radially symmetrical uniform density distribution up to a certain radius:

\begin{equation}
W(kR) = \int_{\mathbb{R}^3} {\rm d}^3{\bf r}\, \frac{3 \Theta(R-r)}{4\pi R^3}\, e^{-i{\bf k}{\bf r}} = 3 \frac{\sin(kR) - (kR)\cos(kR)}{(kR)^3}.
\end{equation}\\

The observed cluster number density is assumed to follow a Poisson statistic, as commonly performed in modern cluster surveys \citep[e.g.,][]{Bocquet2024}. We do not consider the sample variance \citep{Hu2003}, as its impact is negligible given the characteristics of the eRASS1 survey \citep[i.e., a catalog of massive halos and a large volume, see][]{Fumagalli2021}.
We note as $\lambda$ the intensity of the Poisson process (the number density of objects per unit of observable) and $x$ as the vector of the observables. This intensity depends on the background cosmological model. We note with $\Theta$ the set of cosmological parameters used. By definition, the expected number of objects whose observable properties belong to the subsample $\Omega$ of the observable parameter space follows

\begin{equation}
    \label{eq:expected_distribution}
    N_{\{x\in\Omega\}}(\Theta) = \int_{\Omega} \lambda(x|\Theta)\;\mathrm{d}x.
\end{equation}

In the case of eRASS1, combined with the observation of the Legacy Survey DR10 \citep{Kluge2024}, the observable vector incorporates the observed X-ray count rate, $\widehat{C_\mathrm{R}}$; the observed optical richness, $\widehat{\lambda}$; the photometric redshift, $\widehat{z}$; the observed sky position, $\widehat{\mathcal{H}}$; and the reduced tangential shear profile, $\widehat{g_\mathrm{t}}$, for clusters that belong to the overlapping weak lensing surveys used for the mass calibration. The vector of observables consequently reads as

\begin{equation*}
    x = \{\widehat C_\mathrm{R}, \widehat z, \widehat\lambda, \widehat{\mathcal{H}}, \widehat{g_{t}}\}.
\end{equation*}

We can thus rewrite the intensity of the Poisson process as $\lambda(x|\theta)=(\mathrm{d}\,\hat n)/(\mathrm{d}\widehat{C_\mathrm{R}}\mathrm{d}\widehat{\lambda}\mathrm{d}\widehat{g_\mathrm{t}}\mathrm{d}\widehat{z}\mathrm{d}\mathcal{H})$, with 

\begin{equation}
    \label{eq:observed_number_density}
    \frac{\mathrm{d}\,\hat n}{\mathrm{d}\widehat{C_\mathrm{R}}\mathrm{d}\widehat{\lambda}\mathrm{d}\widehat{g_\mathrm{t}}\mathrm{d}\widehat{z}\mathrm{d}\mathcal{H}} = \int\frac{\mathrm{d}\,n}{\mathrm{d}\mathrm{ln} M\mathrm{d}z\mathrm{d}\mathcal{H}}\mathcal{P}(x|M,z)\mathcal{P}(I|x)\, \mathrm{d}\mathrm{ln} M \mathrm{d}z,
\end{equation}

\noindent where $\mathcal{P}(x|M,z)$ is the probability distribution function related to the observables at a given mass and redshift (see Sect.~\ref{sec:scal_rel}), and $\mathcal{P}(I|x)$ is the selection function, describing the expected fraction of objects that we detect for a given set of observable $x$, described in Sect.~\ref{sec:sel_func}.

\subsection{Selection function}
\label{sec:sel_func}

The selection function model $\mathcal{P}(I|x)$ was produced using the state-of-the-art eRASS1 digital twin simulation \citep{Seppi2022, Comparat2020}. The simulation reproduces the AGN and cluster distributions with high accuracy and produces synthetic eRASS1 observations.
A detection probability is assigned to each set of observables using a Gaussian process classifier described in \cite{Clerc2024}. In practice, the observables used to assign the probability are the true redshift $z$, the galactic column density \nh-corrected count rate $C_R$, and the sky position $\mathcal{H}$.
The selection function depends on the sky position through the local background surface brightness, the local hydrogen column density, and the exposure time, which are not uniform across the eRASS1 sky. The selection function enters the statistical inference framework as described in Sect.~\ref{sec:stat_inf}.

\subsection{Scaling relations} 
\label{sec:scal_rel}

The eRASS1 cluster abundance pipeline primarily considers two mass proxies: the X-ray count rates, $C_\mathrm{R}$, and the optical richness, $\lambda$. The scaling relations between these observables and the underlying dark matter halo mass are fitted together with the cosmological parameters. 
The count rate scaling relation is assumed to follow

\begin{equation}
\left\langle \ln \frac{C_\mathrm{R}}{C_\mathrm{R,p}} \bigg| M, z \right\rangle = 
\ln A_\mathrm{X} + 
b_\mathrm{X}(z) \, \ln \frac{M}{M_\mathrm{p}} + e_\mathrm{x}(z) ,
\label{eq:cr_m}
\end{equation}

\noindent where the pivot count rate, pivot mass, and pivot redshift are fixed to $C_\mathrm{R,p} = 0.1$ cts~/~s, $M_\mathrm{p} = 2 \times 10 ^{14} M_\odot$, and $z_\mathrm{p} = 0.35$ by choice. The other terms follow

\begin{equation}
b_\mathrm{X}(z) = B_\mathrm{X} + F_\mathrm{X} \, \ln \frac{1+z}{1+z_\mathrm{p}}
\end{equation}

\noindent and

\begin{equation}
e_\mathrm{x}(z) = D_\mathrm{X} \, \ln \frac{d_\mathrm{L}(z)}{d_\mathrm{L}(z_\mathrm{p})} + E_\mathrm{X} \, \ln \frac{E(z)}{E(z_\mathrm{p})} + G_\mathrm{X} \, \ln \frac{1+z}{1+z_\mathrm{p}}.
\label{eq:red_evol}
\end{equation}

The richness scaling relation has a form similar to 

\begin{equation}
    \label{eq:richness}
    \left\langle \ln \lambda \big| M,z\right\rangle = \ln A_\lambda + b_\lambda(z)\ln\left(\frac{M}{M_\mathrm{p}}\right) + C_\lambda \ln\left(\frac{1+z}{1+z_\mathrm{p}}\right),
\end{equation}

\noindent and the redshift dependence of the mass slope follows

\begin{equation}
    \label{eq:richness_mass_slope}
    b_\lambda(z) = B_\lambda + D_\lambda\ln\left(\frac{1+z}{1+z_\mathrm{p}}\right).
\end{equation}

These scaling relations bridge the X-ray, optical, and shear data. They enter into the statistical inference framework via $\mathcal{P}(x|M,z)$, as described in Sect.~\ref{sec:stat_inf}.

\subsection{Weak lensing calibration}
\label{sec:weak_lensing_calibration}

This section demonstrates that the modeling introduced for the cluster mass calibration in \citetalias{Ghirardini2024} can be used in this work without further modifications. Future X-ray samples will explore lower-mass ranges and will require a deeper understanding of the lensing of dark matter halos in the presence of ULA. In the context of eRASS1 statistical precision, we demonstrate that adopting a generic NFW profile is sufficient. We begin by showing that the weak lensing mass bias model, which characterizes the discrepancy between the true total mass of a cluster and the weak lensing-inferred mass employed in this analysis, is largely insensitive to the ULA mass fraction.
In practice, to accurately predict the mapping between the halo mass and the weak lensing profiles, we employ the shear inferred cluster mass, which results from fitting a reduced shear profile with our custom shear profile model, as proposed by \citet{grandis21b}. However, the shear inferred masses are biased compared to the true total mass of a cluster \citep{Becker2011}. The relation between weak lensing mass and true dark matter halo mass is parametrized with the weak lensing bias and weak lensing scatter. The values of the latter, as well as their uncertainties, are calibrated on the Monte-Carlo realizations of synthetic shear profiles \citep{Grandis2024}. These synthetic shear profiles are based on mass maps from hydrodynamical simulations augmented by the eRASS1 miscentering distribution, by the cluster member contamination results, by the shape and photometric redshift calibration uncertainties of respective weak lensing surveys, and by an uncorrelated large-scale structure noise contribution. We assume that these calibrations will remain valid over the full ULA mass range for small ULA fractions $\oax/(\Omega_\mathrm{cdm}+\oax) \lesssim 0.1$, since simulations do not show evidence of ULA effects on halo density profiles below this threshold \citep{Schwabe20, Vogt23}. As discussed below, even for a dark matter sector comprised solely by ULAs ($\oax/(\Omega_\mathrm{cdm}+\oax) = 1$), the calibrations remain valid for $m_\mathrm{a} \gtrsim 10^{-26}\,\si{\electronvolt}$. Consequently, the bias model remains the same as the one used in \citetalias{Ghirardini2024}.

\begin{figure*}
    \centering
    \includegraphics[width=\textwidth]{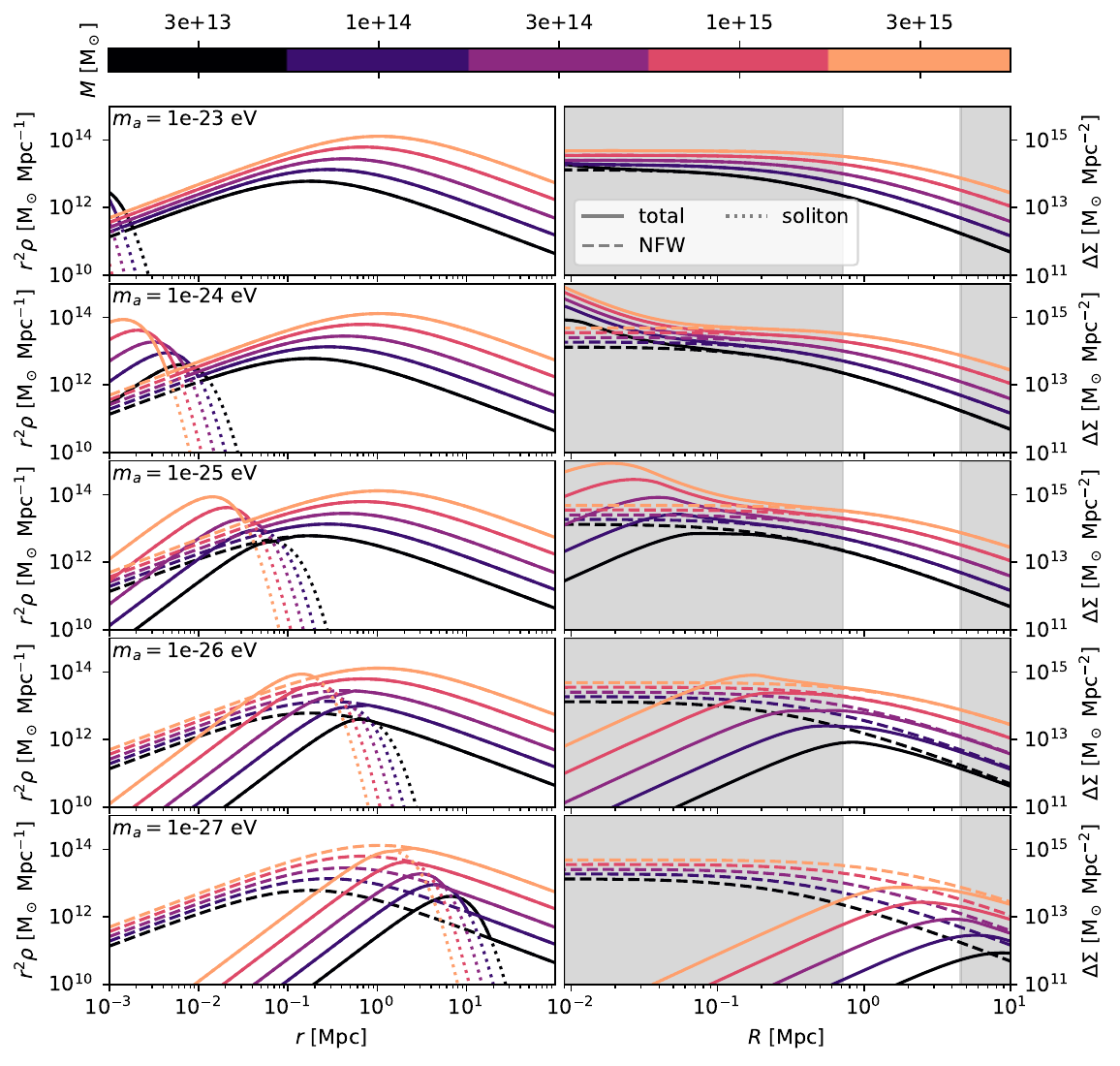}
    \caption{Density profiles (left column) and projected density contrasts (right column) of halos of different mass (color coded) at redshift $z=1$ for different ULA particle masses (rows) in the extreme case of a ULA-only dark matter sector. Full lines show the total profiles, dashed lines show the corresponding cold dark matter profiles (assumed to follow a Navarro-Frenk-White profile), and dotted lines show the soliton core profile. In the right column, we highlight the scales we use for the weak lensing measurement in white.}
    \label{fig:pheno_WL}
\end{figure*}

Furthermore, we show that for an eRASS1-like sample and a cluster abundance-based analysis, the impact of ULAs on the dark matter density profile used in weak lensing mass estimation can be safely neglected. However, for deeper eROSITA surveys, this assumption may break down due to the increased statistical contribution from group-sized halos.
In general, the presence of ULA dark matter impacts the structure of halos, as explored most completely in gravity-only simulations in \citet{May2023}. Those simulations confirm prior phenomenological findings of \citet{schiveb, Marsh2016} that ALPs form a so-called soliton core where quantum degeneracy pressure compensates for the inward pull of gravity in the inner regions of halos. The density profile of the core can be analytically approximated as

\begin{equation}
    \rho_\text{soli}(r) = \frac{ 1.9 \times 10^9 (1+z) \left( \frac{m_\mathrm{a}}{10^{-23} \text{ eV}}\right)^{-2} \left( \frac{r_\text{soli}}{\text{kpc}}\right)^{-4}}  { \left( 1 + 0.091 \left( \frac{r}{r_\text{soli}}\right)^2\right)^8} \text{ M}_\odot \, \text{kpc}^{-3},
\end{equation}

\noindent where $z$ is the halo redshift. The characteristic radius of the soliton cores $r_\text{soli}$ marks the scale at which the density profile transitions from a flat inner part to a steeply declining outer part. Such core profiles for different ULA particle masses $m_\mathrm{a}$, and halo masses $M$ at redshift $z=0.1$ are shown in the left panel of Fig.~\ref{fig:pheno_WL} as dotted lines. Crucial to the phenomenology of halo profiles in ULA cosmology is the halo mass dependence of the core radius, given by \citet{schiveb} as

\begin{equation}
    r_\text{soli} = 1.6 \left( \frac{m_\mathrm{a}}{10^{-22} \text{ eV}}\right)^{-1} (1+z)^{1/2} \left( \frac{\zeta(z)}{\zeta(0)}\right)^{1/6} \left( \frac{M}{10^9 \text{M}_\odot}\right)^{-1/3} \text{ kpc},
\end{equation}

\noindent where $\zeta(z)=(18 \pi^2 + 82(\Omega_\mathrm{m}(z)-1)-39(\Omega_\mathrm{m}(z)-1)^2)/\Omega_\mathrm{m}(z)$, which we evaluate with the present day matter density ${\Omega_\mathrm{m}(z=0)=0.3}$. As shown in Fig.~\ref{fig:pheno_WL}, smaller ULA particle masses lead to larger cores, which are also larger in lower-mass halos. 

In this section and for this demonstration, we use the Navarro-Frenk-White profile for the collisionless part of the profile \citep{Navarro1996} given by

\begin{equation}
    \rho_\text{NFW}(r) = \frac{\rho_0}{\frac{r}{r_\text{s}} \left( 1 + \frac{r}{r_\text{s}}\right)^2},
\end{equation}

\noindent with scale radius $r_\text{s}$ and normalization $\rho_0$. We set these values as follows: For a given halo mass and redshift, we computed the radius $r_{500\text{c}}$ using the definition for spherical overdensity masses and the mean concentration $c_{500\text{c}}$ following the relation by \citet{ragagnin21}. This sets $r_\text{s}=r_{500\text{c}}/c_{500\text{c}}$. Using the profile to compute the mass enclosed in $r_{500\text{c}}$, we set the normalization constant $\rho_0$ in accordance with Eq.~4 in \citet{Navarro1996} adjusted to the overdensity $500$.

\begin{figure*}[h!]
\includegraphics[width=0.495\textwidth, clip]{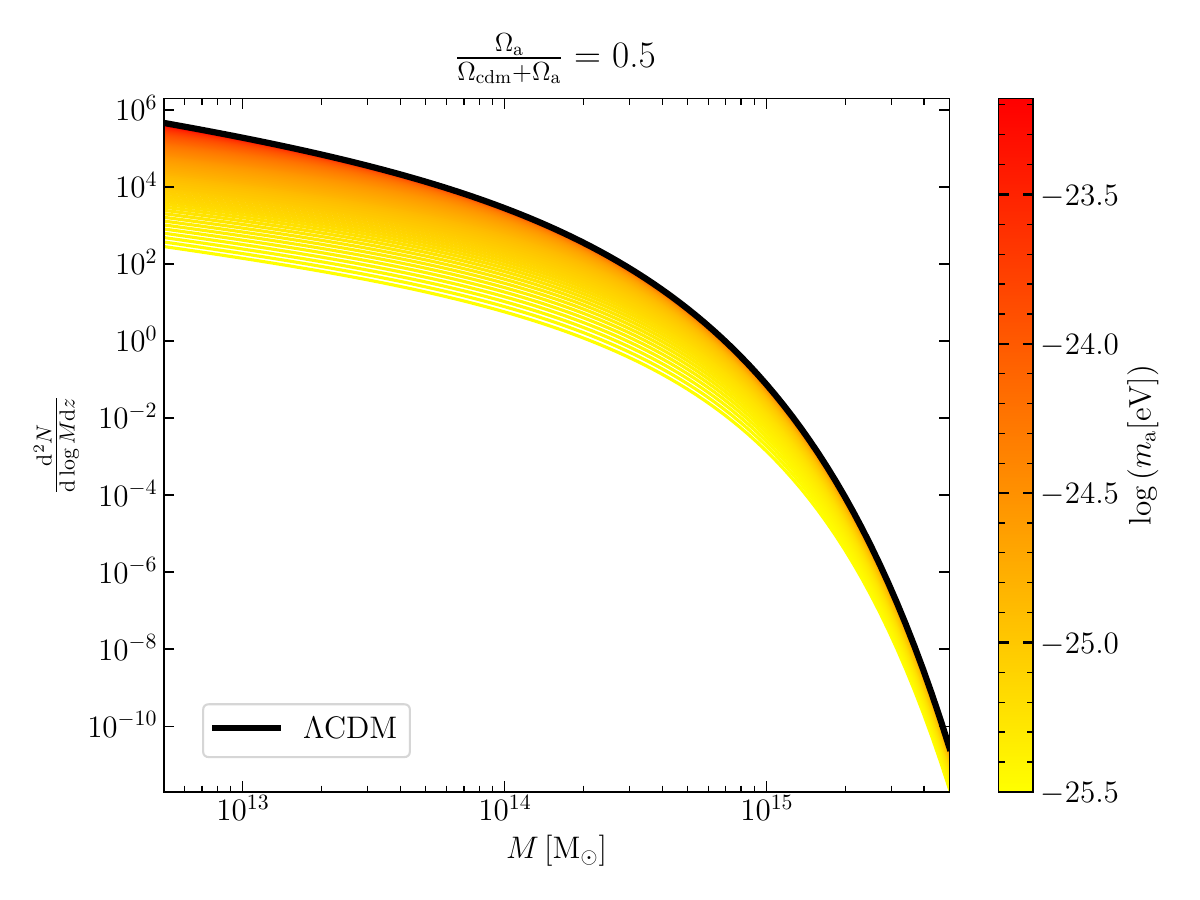}
\includegraphics[width=0.495\textwidth]{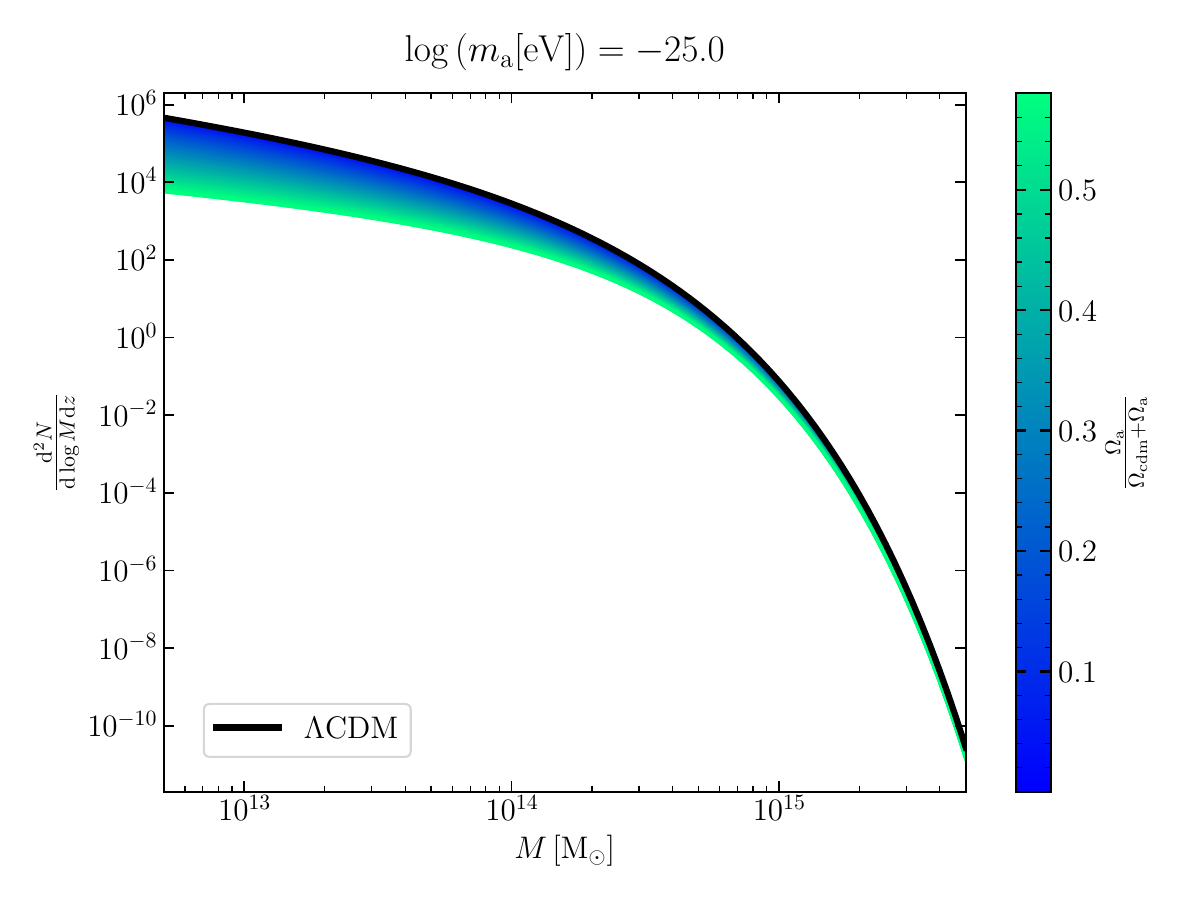} 
\caption{\label{fig:hmf} Halo mass function with \cite{Planck2020} cosmological values ($\Omega_\Lambda \approx 0.7$, $\Omega_\mathrm{m}+\oax \approx 0.3$) at $z=0.1$ with a fixed relative ULA density of $\oax/(\Omega_\mathrm{cdm}+\oax)=0.5$ and varying ULA mass (left) and a fixed ULA mass of $m_{\rm a}=10^{-25} \,\si{\electronvolt}$ and varying ULA abundance (right) compared to the \lcdm HMF (black). Low-mass halo suppression becomes stronger with decreasing ULA mass and increasing ULA abundance. For higher ULA masses of $m_{\rm a} \gtrsim 10^{-24} \,\si{\electronvolt}$, the HMF becomes indistinguishable from the one in a $\Lambda$CDM cosmology, as is the case for $\oax \rightarrow 0$.}
\end{figure*}

We constructed the composite model following the recommendations by \citet{May2023}. The transition radius $r_\text{t}$ from a soliton core to the collisionless profiles is defined by the condition $\rho_\text{NFW}(r_\text{t})=\rho_\text{soli}(r_\text{t})$. If for small ULA masses ($m_\mathrm{a}<10^{-24}$~eV) no $r_\text{t}$ satisfies this condition, we pick $r_\text{t} = \max \left( r^2 \rho_\text{soli} \right)$. If there are two solutions, we picked the larger of the two. The total profile was constructed as

\begin{equation}
    \rho(r) = \begin{cases} \rho_\text{soli}(r) \text{ if } r \leq r_\text{t} \\
    \rho_\text{NFW}(r) \text{ if } r > r_\text{t}.
    \end{cases}
\end{equation}

\citet{May2023} discusses that this model is only qualitatively correct for spherically averaged profiles. As such, it is not suited for our precise weak lensing measurements and can only be used to understand the phenomenological effect of ULA particles on massive halo profiles. In the left column of Fig.~\ref{fig:pheno_WL}, we show the resulting total mass profiles for different ULA masses $m_\mathrm{a}$ and halo masses $M$. 

However, the density profile is not directly observable by the weak gravitational lensing signal, which is sourced by the projected density contrast

\begin{equation}
    \label{eq:excess_surface_mass_density}
    \Delta\Sigma(R) = \langle \Sigma(<R)\rangle - \Sigma(R),
\end{equation}

\noindent where $R$ is the 2D cluster centric distance, and $\Sigma(R) = \int \text{d}z \, \rho(\sqrt{R^2+z^2})$, the projected matter density. The transformation from a 3D density profile to a density contrast is nonlocal, as weak lensing depends on the enclosed mass. Flat density profiles thus provide very little signal, while the effect of central overdensity can be observed well outside of their physical extent. These density contrasts are shown in the right panel of Fig.~\ref{fig:pheno_WL}.

In this extreme case of a ULA-only dark matter sector, two interesting regimes can be observed. For intermediate ULA masses around $m_\mathrm{a} = 10^{-24}$~eV pronounced soliton cores in the innermost part of the halos appear. Nonetheless, this effect is not observable by our weak lensing measurements, which are restricted to scales around $\sim 1$ Mpc (highlighted in white in the right panel of Fig.~\ref{fig:pheno_WL}). Small ULA particle masses $m_\mathrm{a} \lesssim 10^{-26}$~eV are a second interesting regime. For these ULA masses, cores become so extended that they dominate the density profile for group-scale halos. At $m_\mathrm{a} \sim 10^{-27}$~eV, also cluster scale halos with $M<10^{14}$ M$_\odot$ are dominated by cores. This observation can be interpreted in terms of the soliton radius. If it becomes larger than the cluster size, axions no longer cluster in the halo. This is the ULA mass range we refer to as the dark energy regime. In this context, the light axions, because of their flat inner profile shape, provide a vanishing weak lensing signal. In all cases, the excess surface mass density described by equation \ref{eq:excess_surface_mass_density} is mostly unaffected. Furthermore, the effect on $\Delta\Sigma$ is, as expected, more noticeable for low-mass halos ($M\sim 10^{13}M_\odot$). The eRASS1 cosmological sample has a median mass of $M_\mathrm{med}=10^{14.39}M_\odot$ while 90\% of the halos have a mass larger than $10^{14}M_\odot$.
We therefore conclude that even a pure ULA dark matter sector for $m_\mathrm{a} \gtrsim 10^{-26}$~eV does not significantly impact the prediction for the weak lensing signal of group and cluster scale halos, as compared to a Navarro-Frenk-White profile.
Quantitative predictions for the weak lensing signal of groups and clusters would require large-volume hydro-dynamical simulations in ULA cosmology, which are not available to date, and therefore cannot be incorporated in this work. 
These considerations are left for future works.

Finally, in a mixed ULA-cold dark matter model, N-body simulations by \cite{Schwabe20} show that soliton cores are formed for ULA fractions higher than $\oax/(\Omega_\mathrm{m}+\oax) \gtrsim 0.1$. However, below this fraction, these simulations show no evidence of the formation of soliton cores. Below ULA masses of $m_\mathrm{a} \lesssim 10^{-28}$~eV, ULAs are in the dark energy regime and do not contribute to the WL signal (see Sect.~\ref{sec:dark_energy_regime}).

We conclude that the weak lensing mass calibration remains valid throughout the full ULA mass range as long as the ULA fraction remains below the aforementioned threshold. Hence, we use the profile described in \cite{Grandis2024}.

\subsection{Mixture model}
\label{sec:mix_mod}

The eRASS1 cluster catalog contains three classes of objects: galaxy clusters (C), which are of interest, and active galactic nuclei (AGN), as well as background fluctuations misclassified as clusters (NC), the latter two considered contaminants. Our model simultaneously accounts for the cluster counts and the contaminant fractions through the Poisson mixture model, as described in detail in \cite{Kluge2024}. The total density is the sum of the three-component model:

\begin{equation}
    \label{eq:total_intensity}
    \lambda_\mathrm{tot}(x|\Theta) = \lambda_\mathrm{C}(x|\Theta) + \lambda_\mathrm{AGN}(x) + \lambda_\mathrm{NC}(x).
\end{equation}

In terms of the total number of objects in each class, we obtained

\begin{equation}
    \label{eq:def_mix_mod}
    N_\mathrm{tot}(\Theta) = N_\mathrm{C}(\Theta) + f_\mathrm{AGN}N_\mathrm{tot}(\Theta) + f_\mathrm{NC}N_\mathrm{tot}(\Theta),
\end{equation}

\noindent where $f_\mathrm{AGN}$ and $f_\mathrm{NC}$ are the respective fractions of contaminants. Consequently, we can express the number of AGN, false detections, and the total number of objects in the catalog as a function of the cosmology-dependent number of clusters $N_\mathrm{C}$:

\begin{equation}
    \left\{    
     \begin{array}{ll}
         N_\mathrm{tot}(\Theta) = (1/(1-f_\mathrm{AGN}-f_\mathrm{NC}))N_\mathrm{C}(\Theta) \\
         N_\mathrm{AGN}(\Theta) = (f_\mathrm{AGN}/(1-f_\mathrm{AGN}-f_\mathrm{C}))N_\mathrm{C}(\Theta) \\
         N_\mathrm{NC}(\Theta) = (f_\mathrm{NC}/(1-f_\mathrm{AGN}-f_\mathrm{NC}))N_\mathrm{C}(\Theta) \\ 
    \end{array}
    \right. .
    \label{eq:total_number}
\end{equation}

\begin{figure}[h!]
    \centering
    \includegraphics[width=0.5\textwidth]{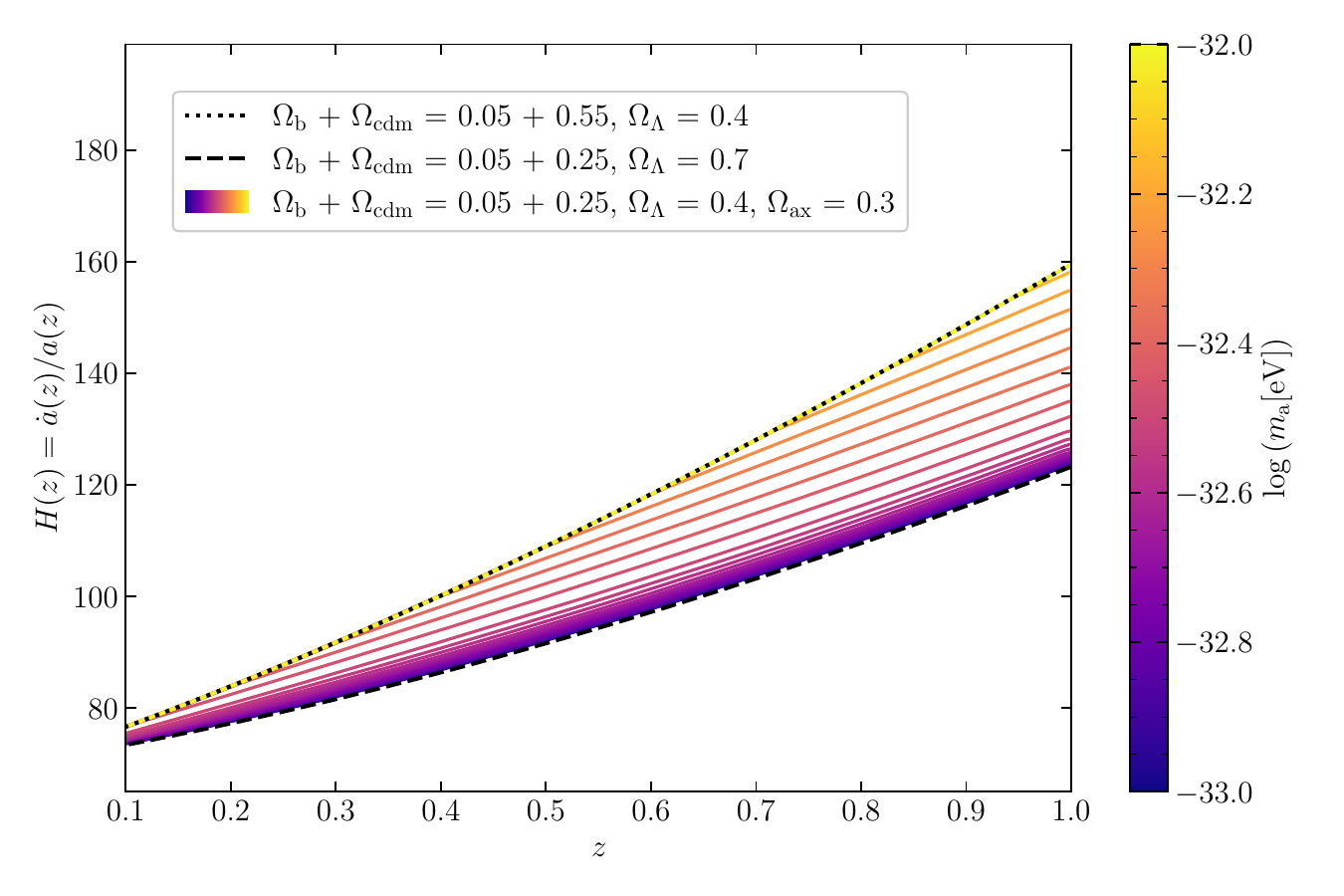}
    \caption{Hubble rate $H(z)=\dot a(z)/a(z)$ as a function of redshift. The dotted black line assumes a flat \lcdm cosmology with $\Omega_\mathrm{m}=0.6$ and $\Omega_\Lambda=0.4$. The dashed black line assumes a flat \lcdm cosmology with $\Omega_\mathrm{m}=0.3$ and $\Omega_\Lambda=0.7$. The colored lines represent Hubble rates for a ULA cosmology with fixed relative densities $\oax=0.3$, $\Omega_\mathrm{m}=0.3$, and $\Omega_\Lambda=0.4$. The colored Hubble rates differ only in the ULA mass, as shown by the color bar. This plot demonstrates that ULAs undergo a domain change in the low-mass regime between $\logm=-32$ and $\logm=-33$. On the low-mass end, the Hubble rate approaches that of a flat \ LCDM cosmology but explains parts of the dark energy component. In the modifications to the pipeline used in \citetalias{Ghirardini2024}, we use this accurate Hubble rate extracted from \texttt{axionCAMB} in the distance measure computations.}
    \label{fig:hubble}
\end{figure}

Starting from the formalism above, we can write the total number of objects as 

\begin{equation}
    \label{eq:total_obs}
    N_\mathrm{tot}(\Theta) = \frac{1}{1-f_\mathrm{AGN}-f_\mathrm{NC}}\int_x \lambda_\mathrm{C}(x|\Theta) \;\mathrm{d}x,
\end{equation}

\noindent and that the number density of contaminants follows $\lambda_\mathrm{AGN}(x|\theta) = N_\mathrm{AGN}(\Theta)\mathcal{P}_\mathrm{AGN}(x)$, and $\lambda_\mathrm{NC}(x|\theta) = N_\mathrm{NC}(\Theta)\mathcal{P}_\mathrm{NC}(x)$. In this equation, $\mathcal{P}$ describes the probability distribution function of the respective object, depending on the observables. Finally, the likelihood  becomes

\begin{equation}
    \label{eq:likelihood_cont}
    \begin{array}{ll}
    \ln \mathcal{L}(\Theta) = & \displaystyle\sum_i \ln\; \Bigg(\lambda_\mathrm{C}(x_i|\Theta) \\
   &+ N_\mathrm{AGN}(\Theta)\mathcal{P}_\mathrm{AGN}(x_i) + N_\mathrm{NC}(\Theta)\mathcal{P}_\mathrm{NC}(x_i) \Bigg)  \\
    &- \,\displaystyle\frac{1}{1-f_\mathrm{AGN}-f_\mathrm{NC}}\int_x \lambda_\mathrm{C}(x|\Theta) \;\mathrm{d}x\; .
    \end{array}
\end{equation}

The contaminant fractions $f_\mathrm{AGN}$ and $f_\mathrm{NC}$ are fit for during the parameter inference.

\begin{figure}[h!]
\includegraphics[width=0.5\textwidth]{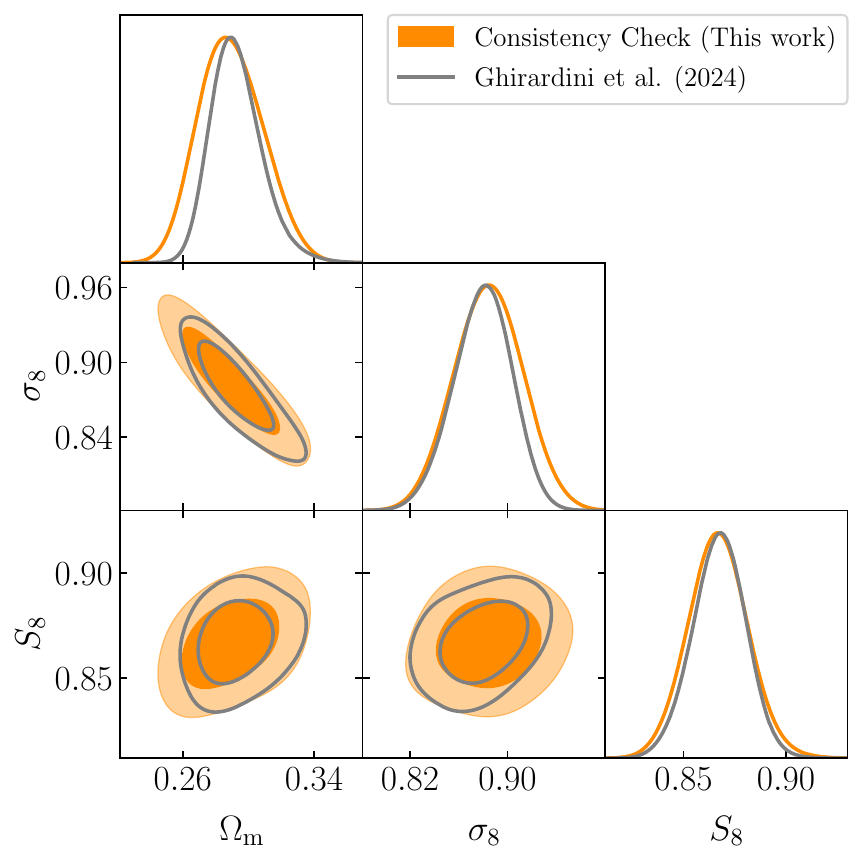}
\centering
\caption{Consistency check: Posteriors for the cosmological parameters $\Omega_\mathrm{m}$, $\sigma_8$, and $S_8$ in the $\Lambda$CDM ($m_{\rm a} = 10^{-12}$~eV and $\oax=10^{-9}$). These values were chosen arbitrarily such that they satisfy $m_{\rm a} \gg 10^{-18}$~eV and $\oax \ll 10^{-3}$. The results are consistent with \citetalias{Ghirardini2024}. The inner and outer contours represent the 68\% and 95\% contour levels, respectively.}
\label{fig:FigTEMPA1}
\end{figure}

\subsection{Integration of ultralight axions}
\label{sec:integration_of_ulas}

Constraining the ULA fraction requires us to predict the massive dark matter halo number density in the corresponding model.
We introduce ULAs in the cluster abundance cosmological pipeline by implementing a modified version of the power spectrum used to produce the HMF \citep{Diehl2021}. This framework provides us with the proper transfer function and growth of structure models.
In practice, we use {\texttt {axionCAMB}} \citep{Grin2022axionCAMB}, a modified version of {\texttt {CAMB}} that adds the evolution of primordial density fluctuations caused by an axion field to the Boltzmann solver. The effect of ULAs on the HMF is illustrated in Fig.~\ref{fig:hmf} for different ULA masses $m_\mathrm{a}$ and ULA relic densities \oax.
We use {\texttt {axionCAMB}} to modify the Hubble rate as a function of redshift $H(z)$ for a universe with dark matter composed of ULAs. Using an updated Hubble rate ensures that we properly account for the redshift dependence of the Hubble rate on the effective properties of ULAs (dark energy or dark matter behavior in the Hubble rate) when computing the distances between background galaxies and the lens (or the observer) needed for the weak gravitational lensing calibration of the scaling laws a described in Sect.~\ref{sec:weak_lensing_calibration}. Fig.~\ref{fig:hubble} shows the influence of ULA mass on the Hubble rate in a cosmology with a high ULA abundance as dark matter.
We also use {\texttt {axionCAMB}} to compute the matter power spectrum.  We then compute the r.m.s. density fluctuations (Eq.~\ref{eq:rms_density}) and use them as input to the multiplicity function $f(\sigma)$ from \cite{Tinker2008} to compute the halo number density. 

\begin{figure}[h!]
\includegraphics[width=0.5\textwidth]{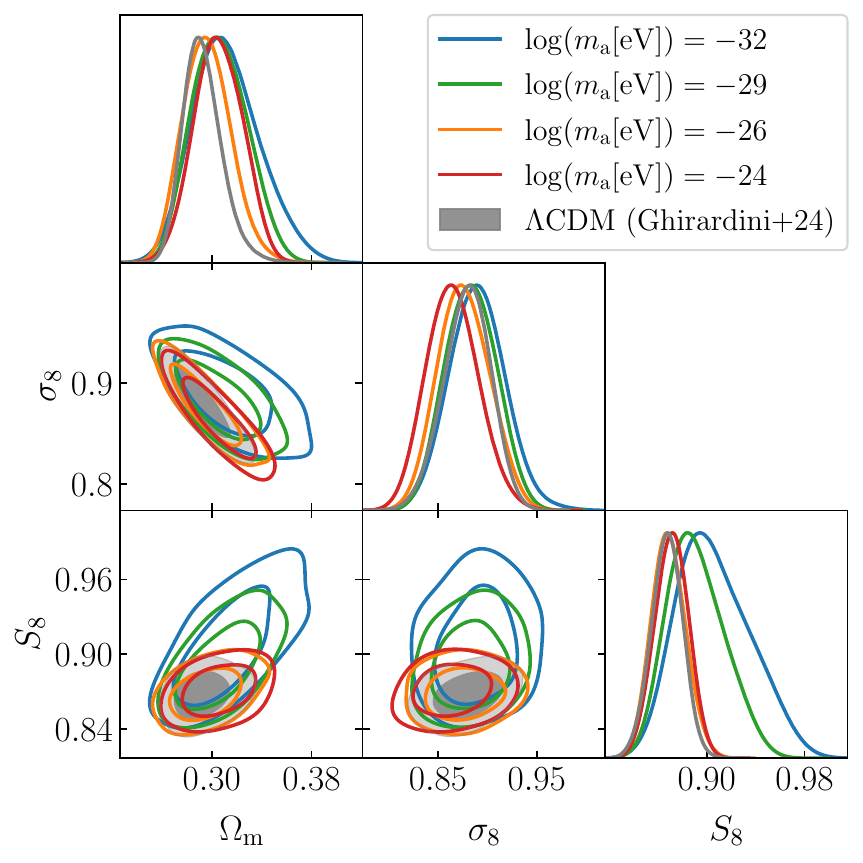}
\centering
\caption{Posterior distributions for \om, \sigmaeight, and \Seight in a subset of ULA mass bins. We find consistent cosmological constraints across all mass bins within 1$\sigma$ (except for the excluded $m_\mathrm{a}=10^{-25}$~eV mass bin). The results are furthermore compatible with \citetalias{Ghirardini2024}. The inner and outer contours represent the 68\% and 95\% contour levels, respectively.}
\label{fig:cosmo_constraints_full_range}
\end{figure}

Consequently, in this work, we choose to model the effect of ULAs solely by implementing the power spectrum produced by the corresponding ULA model in the HMF obtained by \cite{Tinker2008}. The impact of ULAs is thus obtained through the transfer function and the growth of structure used to compute the r.m.s. density fluctuations in equation \ref{eq:rms_density}. However, accounting for ULAs also requires recalculating the critical overdensity for halo collapse, given that they do not form structures below the Jeans mass.
Different studies in the literature have provided fitting functions to model this effect, with smoothed $k$ window functions \citep{Bohr2021}, or analytical formulas \citep{Dome2025}. However, these methods have not been tested in the framework of cosmological parameter inference. Additionally, most of the correcting effects are significant at masses much below the mass range of clusters below the detection limits of the galaxy cluster catalogs of the eROSITA All-Sky Survey ($\sim 10^{10} M_\odot$ for the typical ranges of the corrections, compared with $\sim 10^{14} M_\odot$ for a typical cluster detected in eRASS1). Consequently, it is not necessary to apply these corrections for the eRASS1 sample. We note that better modeling will be required to extend the analysis to a lower-mass range below $M\lesssim10^{13}M_\odot$.

Below an ULA mass of $10^{-28}\,\si{\electronvolt}$, we exclude ULAs from the matter density in the HMF formalism to account for the fact that their de Broglie wavelength exceeds the Jeans length and ULAs are not expected to contribute to halo formation in this regime. For ULA masses of $\logm \lesssim -32.5$, the de Broglie wavelength of the ULA BEC exceeds the size of the universe, making ULAs indistinguishable from a cosmological constant $\Lambda$. As such, a model is indistinguishable from \lcdm, we exclude this regime from our analysis.

For ULA masses $-32.5 \lesssim \logm \lesssim -31.5 $, despite not contributing to halo formation, ULAs affect the Hubble rate, which in turn influences the distance measurements relevant for the weak lensing calibration. Above $\logm \sim -32.5$, only objects at redshifts higher than $z=0.8$ (maximum redshift for galaxy clusters in the eRASS1 cosmology sample) are affected by the regime change in the Hubble rate. This involves the galaxies used for the gravitational weak lensing calibration of the scaling relations (see Sect.~\ref{sec:weak_lensing_calibration} for details) that extend up to $z \lesssim 6$ in redshift space. We consider this effect by computing the distances based on the ULA Hubble rate extracted from \texttt{axionCAMB} in this regime. All ULA mass bins above the lowest-mass bin are not affected by the regime change in the Hubble rate, i.e., ULA masses above $\logm \lesssim -31.5$ (see Fig.~\ref{fig:regimechange} for further details). Fig.~\ref{fig:hubble} illustrates how the regime affects the Hubble rate in a cosmology with a high ULA abundance as a function of ULA mass (color coded).

\section{Results and discussion}
\label{sec:results}

\begin{figure}[t]
    \centering
    \includegraphics[width=\linewidth]{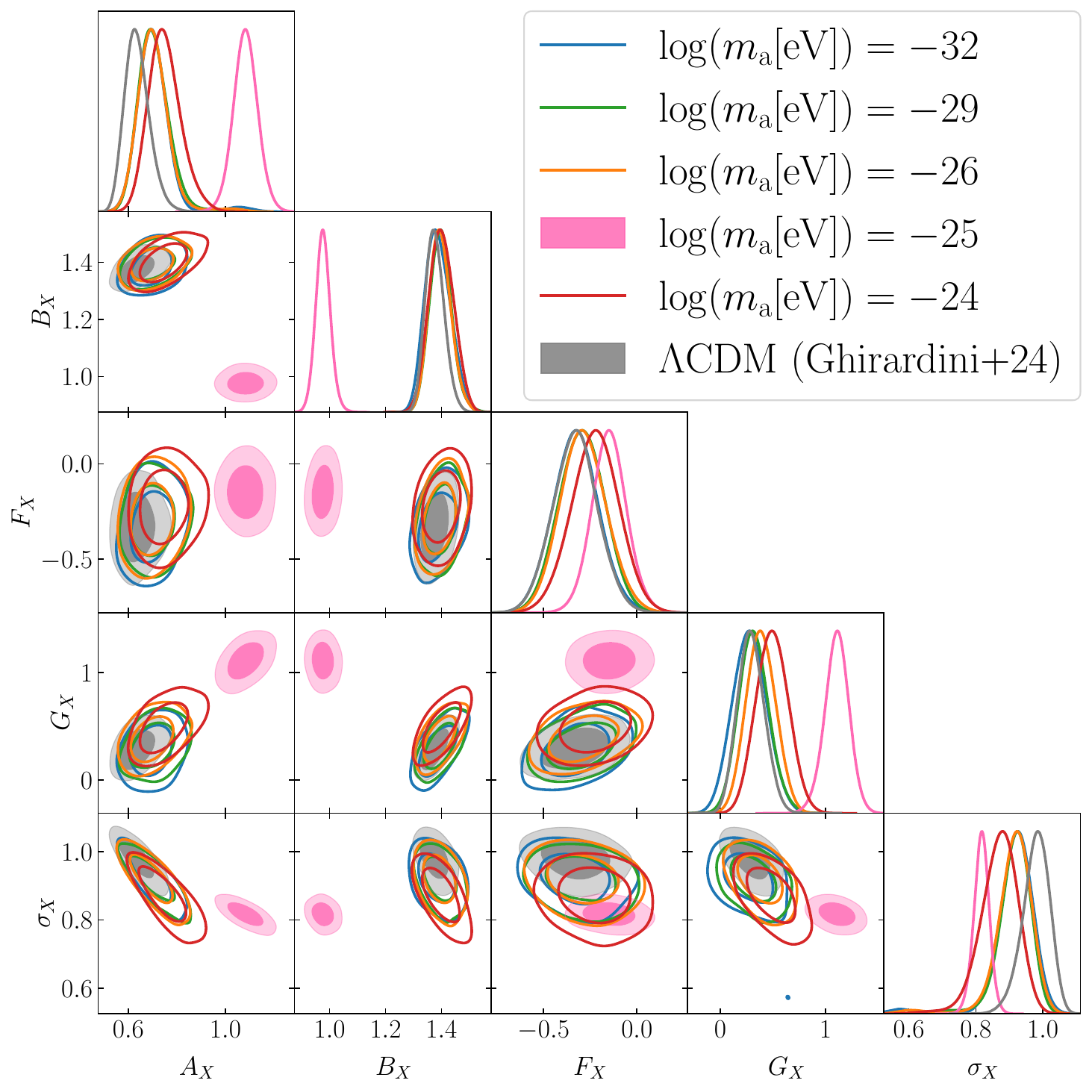}
    
    \vspace{1em} 
    
    \includegraphics[width=\linewidth]{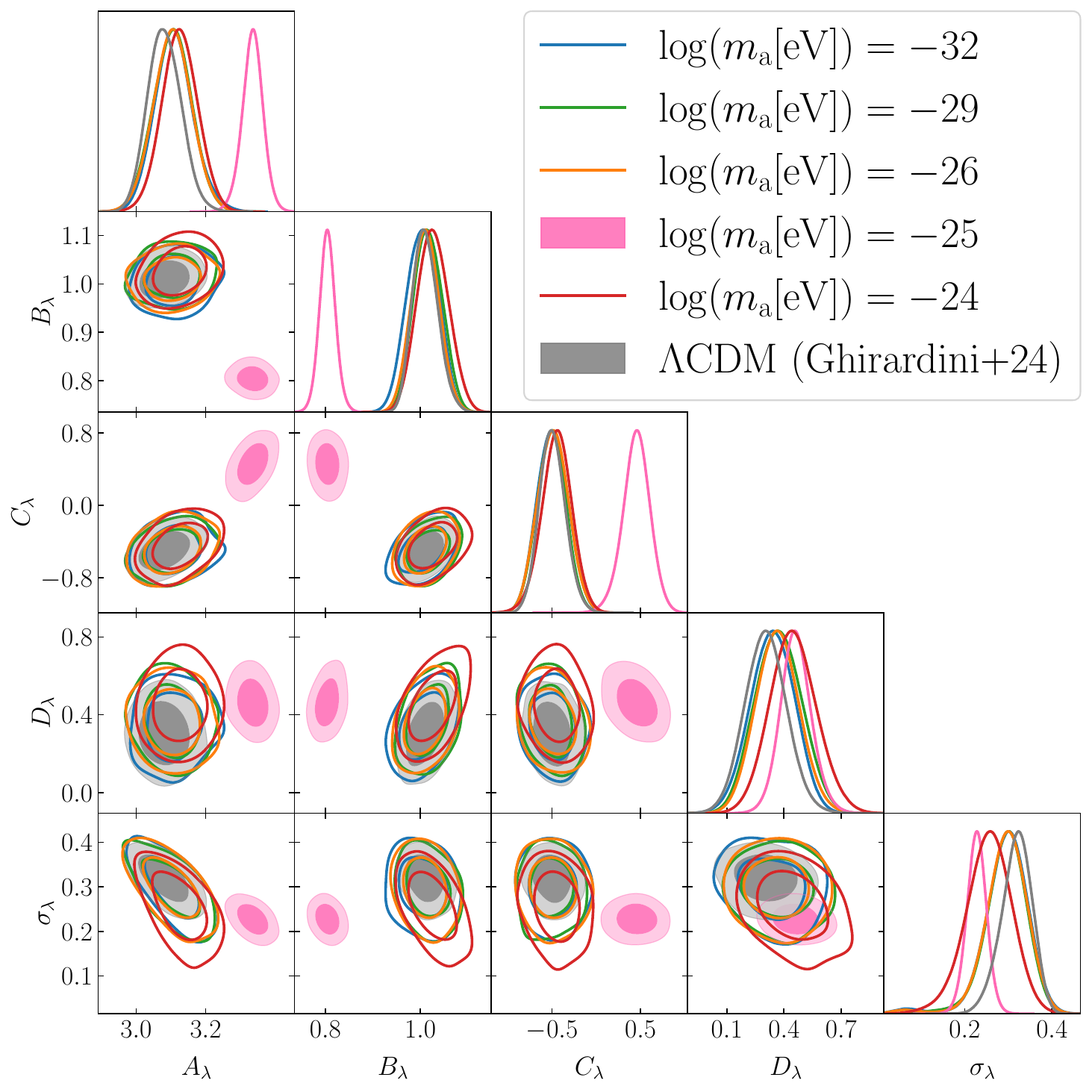}
    \caption{Posterior distributions in a representative subset of ULA mass bins for the X-ray scaling relations (top) and the optical scaling relations (bottom). The large discrepancies in the astrophysical scaling relation model parameters between the $m_\mathrm{a}=10^{-25}$~eV mass bin on the one hand and \citetalias{Ghirardini2024} and the remaining mass bins on the other hand indicates an unphysical posterior distribution in the $m_\mathrm{a}=10^{-25}$~eV ULA mass bin. Consequently, the concerned mass bin is excluded from the analysis.}
    \label{fig:sl_discrepancies}
\end{figure}

\begin{figure*}[h!]
    \centering
    \includegraphics[width=0.495\textwidth]{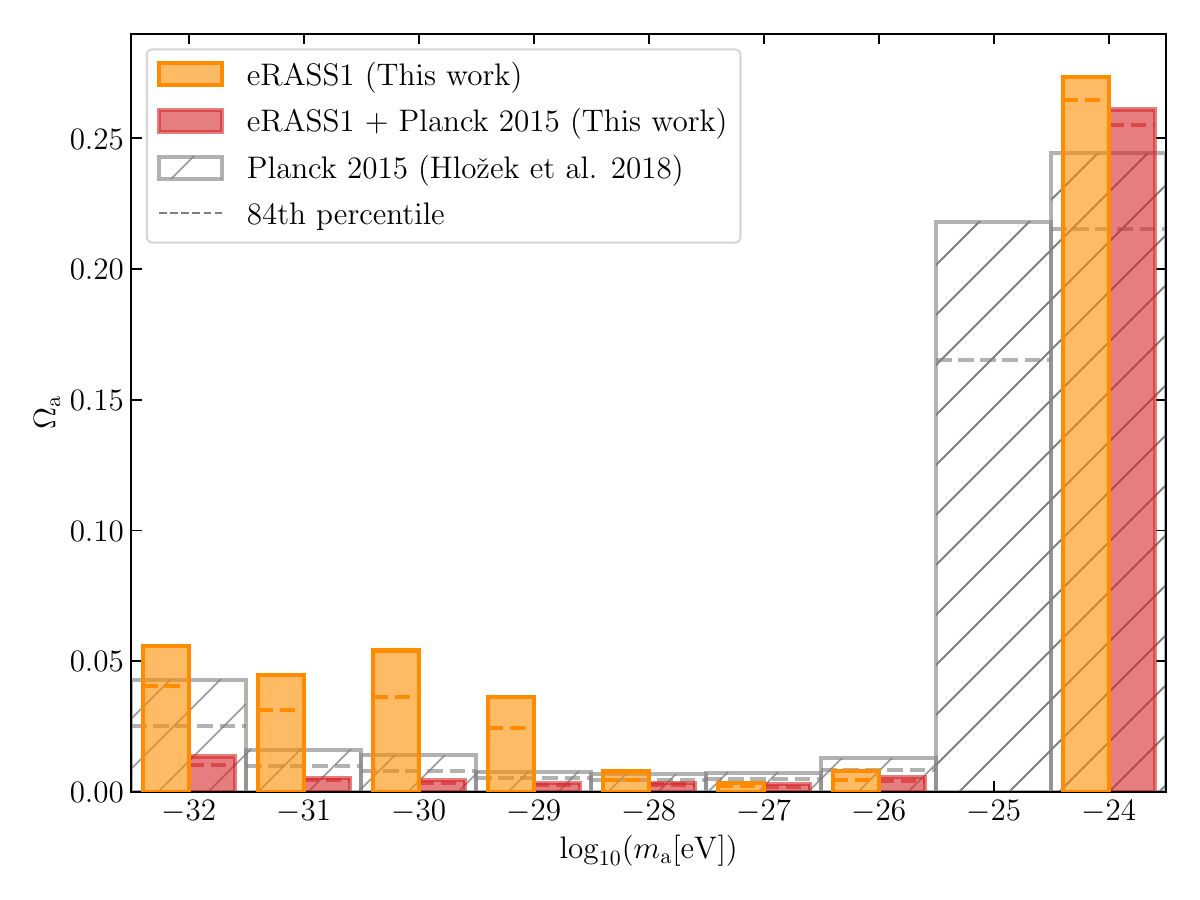}
    \includegraphics[width=0.495\textwidth]{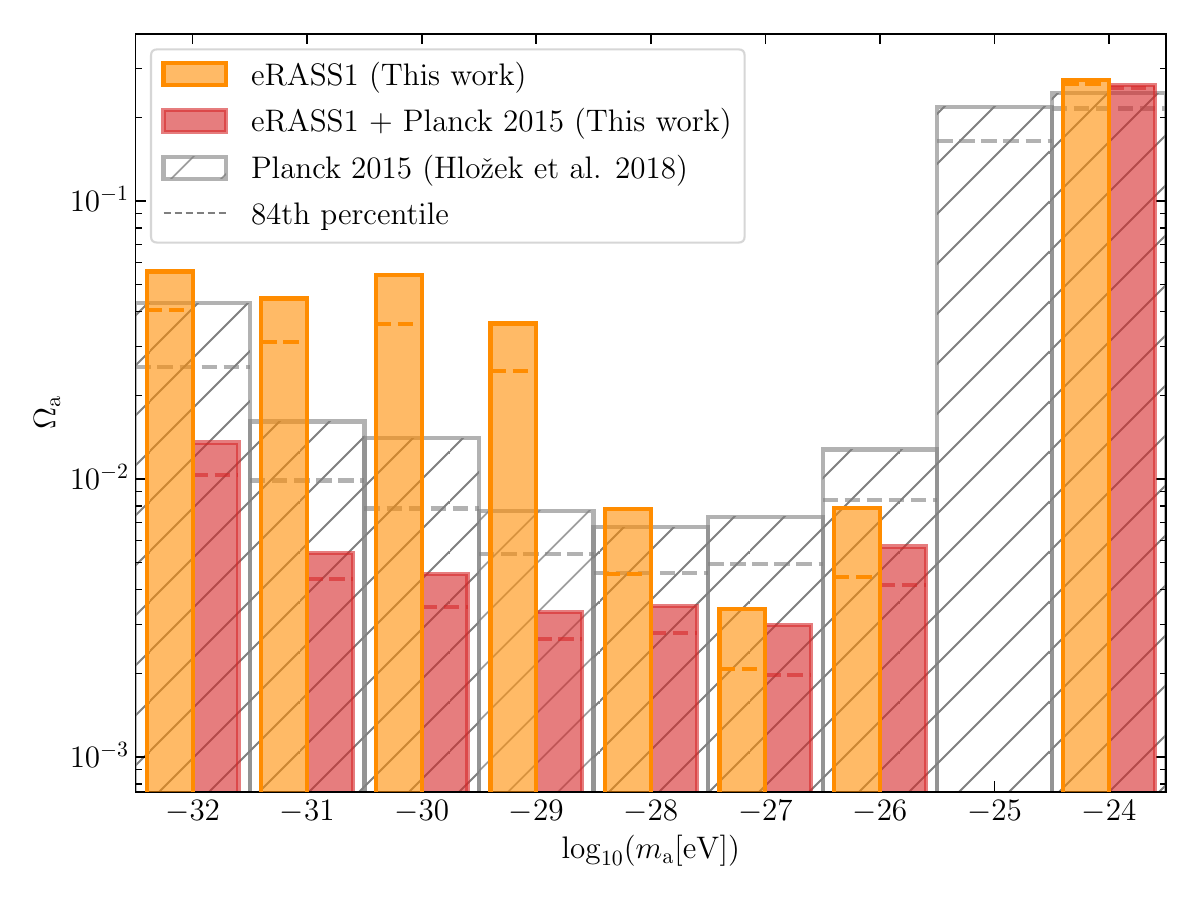}
    \caption{Upper bounds on the relative ULA density $\oax$ in each logarithmic ULA mass bin. The left plot shows the upper bounds on \oax on a linear scale, and the right plot shows the upper bounds on a logarithmic scale. The bar height indicates the 95\% confidence interval. The orange bars indicate the bounds obtained by eRASS1 galaxy clusters only, and the gray hatched bars show the bounds obtained from \planck 2015 CMB data by \cite{Hlozek2018}. The red bars show the bounds obtained from combining the bounds from eRASS1 galaxy clusters with those from \planck 2015 CMB data.}
    \label{fig:constraints_2D}
\end{figure*}

In this section, we present the results of the Bayesian inference fitting of cosmological and ULA parameters through galaxy cluster number counts. We used the X-ray count rate, the optical richness, and (where available) the reduced tangential shear profiles given in the eRASS1 galaxy cluster cosmology catalog.

\subsection{Consistency checks}
\label{sec:consistency_check}

We implement several changes to the base cosmological pipeline of \citetalias{Ghirardini2024} as described in detail in Sect.~\ref{sec:methodology}. To ensure the adaptations do not introduce inconsistencies, we reproduce the main cosmological results in the new ultralight axion framework by choosing fixed ULA parameters, which are indistinguishable from a \lcdm cosmology. This limit can be achieved either by setting the ULA relic density to $\oax=0$ or for values $\logm \gtrsim -18$ (see Fig.~\ref{fig:matterpowerspectrum} and Fig.~\ref{fig:hmf}). The code \texttt{axionCAMB} only allows positive values for $\oax$, so we chose $\oax=10^{-9}$, several orders of magnitude below any other cosmological fluid component in the relevant redshift range. We additionally fix the ULA mass to $\logm = -12$. The matter power spectrum and hence $\sigma(M, z)$ with this choice of ULA parameters is in perfect agreement with the \lcdm quantities. We checked that running the adapted pipeline leads to almost identical results as compared to the ones obtained in the \lcdm case shown in \citetalias{Ghirardini2024}. A subset of the obtained cosmological parameters is presented in Fig.~\ref{fig:FigTEMPA1}. We conclude that the applied changes did not introduce inconsistencies with the main cosmological results from eRASS1.

\begin{table}
\caption{Cosmological parameters and priors.}              
\label{table:priors}      
\centering    
\scalebox{0.95}{
\begin{tabular}{l | l c l}         
\hline\hline                       
 & Definition & Unit & Priors \\    
\hline                                   
    $\Omega_\mathrm{m}$ & Matter density parameter & - & $\mathcal{U}(0.05, 0.95)$ \\   
    $\oax$ & ULAs density parameter & - & $\mathcal{U}(0, 1)$ \\
    $\log_\mathrm{10}(A_\mathrm{S})$ & Amplitude of the PS & -     & $\mathcal{U}(-10, -8)$ \\
    $H_\mathrm{0}$ & Hubble parameter & $\frac{\rm km}{{\rm s}{~\rm Mpc}}$      & $\mathcal{N}(67.77, 0.6)$ \\
    $\Omega_{\mathrm{b}}$ & Baryon density parameter  & -     & $\mathcal{U}(0.046, 0.052)$ \\
    $n_s$ & Spectral index of the PS & -  & $\mathcal{U}(0.92, 1.0)$ \\
\hline                       
\end{tabular}
}
\tablefoot{These priors are used for all the ULA mass bins. For the full set of priors used for the other parameters, including scaling relations and nuisance parameters, see \citetalias{Ghirardini2024}. With $\mathcal{U}({\rm min}, {\rm max})$ we indicate a uniform distribution between `min' and `max'. With $\mathcal{N}(\mu, \sigma)$ we indicate a normal distribution centered on $\mu$ and with standard deviation $\sigma$.}
\end{table}

\begin{table*}[h]
    \caption{Upper bounds on the ULA relic density ($\oax$) and $\oax h^2$ at the 95\% confidence level in each constraining mass bin for eRASS1 cluster number counts only and when combined with \planck 2015 CMB data.}
    \label{tab:mergedbounds}
    \centering
    \begin{tabular}{c c c c c}
    \hline
    \hline
    $m_\mathrm{a}$ & $\oax$ & $\oax$ & $\oax h^2$ & $\oax h^2$ \\
    & (eRASS1) & (eRASS1 + \planck CMB) & (eRASS1) & (eRASS1 + \planck CMB) \\
    \hline
        $10^{-26}$ eV & $<0.00790$ & $<0.00573$ & $<0.00361$ & $<0.00263$ \\
        $10^{-27}$ eV & $<0.00341$ & $<0.00299$ & $<0.00156$ & $<0.00135$ \\
        $10^{-28}$ eV & $<0.00780$ & $<0.00351$ & $<0.00353$ & $<0.00156$ \\
        $10^{-29}$ eV & $<0.03627$ & $<0.00333$ & $<0.01672$ & $<0.00137$ \\
        $10^{-30}$ eV & $<0.05409$ & $<0.00457$ & $<0.02492$ & $<0.00181$ \\
        $10^{-31}$ eV & $<0.04463$ & $<0.00543$ & $<0.02063$ & $<0.00227$ \\
        $10^{-32}$ eV & $<0.05580$ & $<0.01354$ & $<0.02583$ & $<0.00592$ \\
    \hline
    \hline
    \end{tabular}
\end{table*}

\subsection{Cosmological constraints}
\label{sec:cosmo_constraints}

Following the approach developed by \cite{Hlozek2018} and \cite{Rogers2023}, we perform a binned analysis in the logarithmic ULA mass \logm. We consider a total of nine bins, from $\logm=-24$ in the highest-mass bin to $\logm=-32$. Independent cosmological and scaling relation parameters are considered within each ULA mass bin. The analysis is not sensitive to how the ULA mass is treated within a mass bin, as drawing the ULA mass randomly within a bin or fixing it to the central bin value does not influence the posterior distribution.

We assumed a flat (\lcdm+ ULA) cosmology with massless neutrinos to get a conservative estimate of the effect of ULAs. We leave the matter energy density fraction at redshift $z=0$, $\Omega_{\rm m}$, the baryon energy density fraction at redshift $z=0$, $\Omega_{\rm b}$, the scalar spectral index, $n_{\rm s}$, the normalization of the matter power spectrum, $\log{(A_{\rm s})}$, and the Hubble constant, $H_0$, free. We leave the same parameters free for the scaling relations as done in \citetalias{Ghirardini2024}.  The priors on the cosmological parameters are shown in Table~\ref{table:priors}. 

For all mass bins except the one with $m_\mathrm{} = 10^{-25}$~eV, we find no deviations from the \lcdm cosmological posteriors obtained by \citetalias{Ghirardini2024}, and the posteriors are in good agreement between the different ULA mass bins. We show the cosmological posterior distributions for the cosmological parameters $\Omega_\mathrm{m}$, $\sigma_8$, and $S_8$ for a subset of ULA mass bins covering the full mass range in Fig.~\ref{fig:cosmo_constraints_full_range}.

In the mass bin around $m_\mathrm{a} = 10^{-25}$~eV (i.e., covering the range $10^{-25.5}\text{~eV} \leq m_\mathrm{a} \leq 10^{-24.5}$~eV), we find significant discrepancies compared to \citetalias{Ghirardini2024} in the X-ray and optical scaling relation parameters (e.g., the discrepancy between $A_\mathrm{X}$ and $G_\mathrm{X}$ is $10.9\sigma$, the discrepancy between $A_\lambda$ and $C_\lambda$ is $8.0\sigma$). Figure~\ref{fig:sl_discrepancies} shows the posterior distributions of both the X-ray and optical scaling relations, comparing a representative selection of ULA mass bins with \citetalias{Ghirardini2024}. The scaling relations, together with the gravitational weak lensing calibration, translate the observed count-rates and optical richnesses to the total masses of the clusters in the eRASS1 sample. Together with the contamination model and the selection function, these relations include all astrophysical modeling between observation and theory. As demonstrated in Sect.~\ref{sec:weak_lensing_calibration}, ULAs do not alter the weak lensing signal compared to the \lcdm scenario in the $m_\mathrm{a} = 10^{-25}$~eV mass bin. Consequently, the astrophysical modeling should be independent from the abundance of ULAs with a mass around $m_\mathrm{a}=10^{-25}$~eV, i.e., the scaling relation parameter posteriors should be comparable with the ones reported by \citetalias{Ghirardini2024}. However, in contrast to all other ULA mass bins, this is not the case for the $m_\mathrm{a}=10^{-25}$~eV bin (see Fig.~\ref{fig:sl_discrepancies}). Thus, the cosmological parameters and the ULA relic density in this bin are not reliable, which is why we opt to exclude the $m_\mathrm{a}=10^{-25}$~eV mass bin from the analysis.

\begin{figure}[h!]
    \centering
    \includegraphics[width=0.5\textwidth]{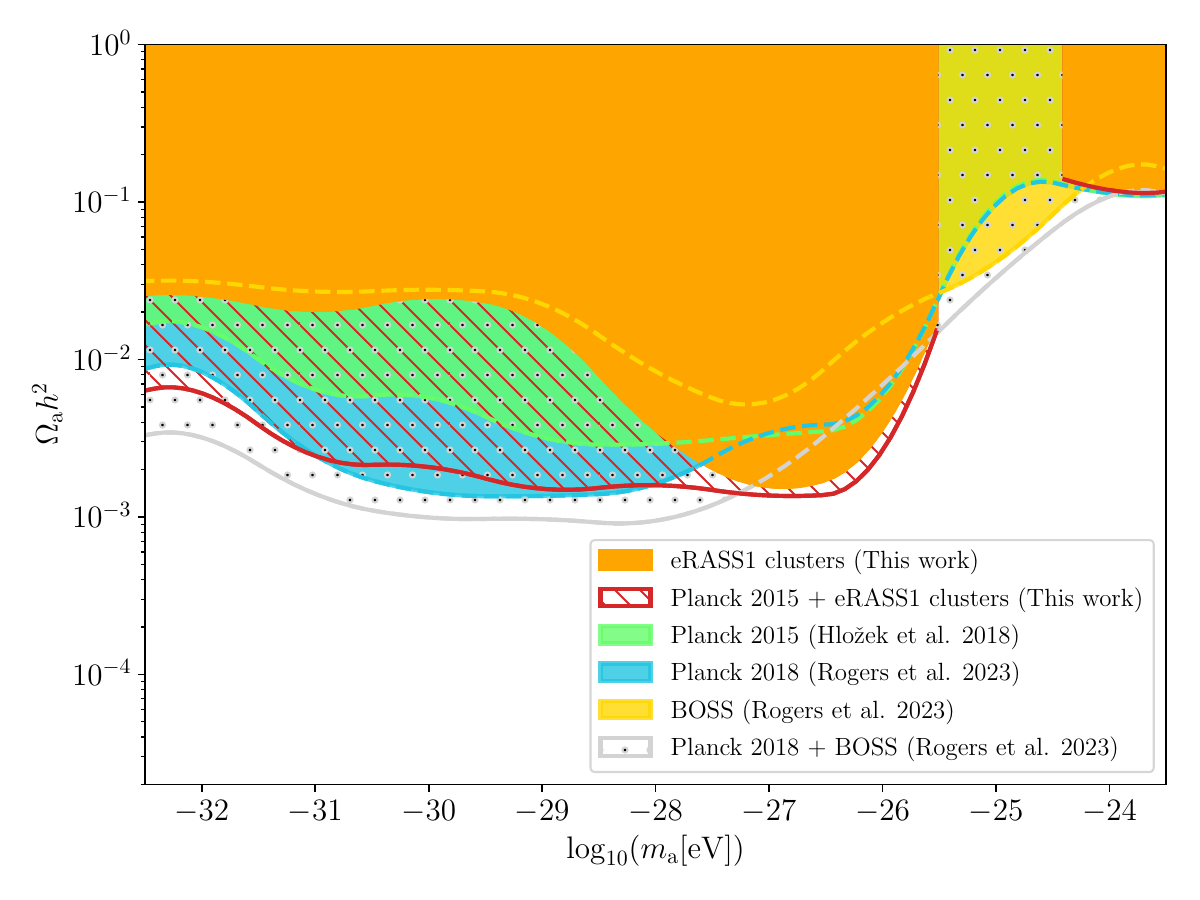}
    \caption{Exclusion regions (95\% confidence level) in the $\logm-\oax h^2$ space as constrained by different probes. Colored regions represent the part of the parameter space where one ULA species can be excluded at 95\% confidence level, whereas white regions represent the unconstrained parameter space. eRASS1 galaxy cluster number counts (orange) exclude a new region in the parameter space between $10^{-27.5}~\si{\electronvolt} \lesssim m_\mathrm{a} \lesssim 10^{-26}~\si{\electronvolt}$, corresponding to galaxy cluster scales. Combining eRASS1 galaxy cluster number counts with the \planck 2015 CMB data (red) further improves the eRASS1 bounds on the ULA relic density. For ULA masses below $m_\mathrm{a} \lesssim 10^{-28}~\si{\electronvolt}$, \planck 2018 CMB data yields the tightest bounds. Above $m_\mathrm{a} \gtrsim 10^{-25}~\si{\electronvolt}$, galaxy clustering data from BOSS yields the tightest constraints. The curves are interpolated between the central bin values, which represent the 95\% confidence levels in the respective bins. In the $m_\mathrm{a}=10^{-25}$~eV mass bin, the \planck upper confidence level has been assumed as a reasonable choice for the interpolation (for details on that bin, see Sect.~\ref{sec:cosmo_constraints}).}
    \label{fig:exclusionregion}
\end{figure}

Furthermore, we find that the posterior distribution is highly sensitive to the choice of priors, indicating a lack of statistical robustness in the $m_\mathrm{a} = 10^{-25}$~eV ULA mass bin. However, we report no constraining power on the ULA relic density in this bin for prior choices comparable to those used by \citetalias{Ghirardini2024}, similar to the results for the $m_\mathrm{a} = 10^{-24}$~eV mass bin. Consequently, the $m_\mathrm{a} = 10^{-25}$~eV mass bin does not contribute to constraining the ULA fraction. Possible reasons for the reported discrepancies include a breakdown of the HMF fitting function or the modeling of the shear signal for this specific ULA mass range. A detailed understanding of this effect will require further investigation beyond the scope of this work. In the figures with an interpolated curve between the bin values (Figs.~\ref{fig:exclusionregion},~\ref{fig:forecats_omega_a}), we assume the \planck upper confidence level \citep{Hlozek2018} as a reasonable reference point for a bin with lacking constraining power for interpolation purposes only.

\subsection{Constraints on the ULA parameter space}
\label{sec:ula_constraints}

We found an exclusion region for ULAs by performing a binned analysis, providing upper bounds on the relic density within logarithmic mass bins. Since ULAs form BECs with characteristic extensions given by their thermal de Broglie wavelengths $\lambda_\mathrm{dB} \sim 1/m_\mathrm{a}$, ULAs with different masses have imprints on different corresponding scales. Ultralight axions with masses of order $\mathcal{O}(10^{-22}\mathrm{~eV})$ have de Broglie wavelengths of order $\mathcal{O}(\mathrm{kpc})$, while ULAs with masses of order $\mathcal{O}(10^{-33}\mathrm{~eV})$ have de Broglie wavelengths comparable to the size of the universe. As galaxy clusters are probing the $\mathcal{O}(\mathrm{Mpc})$ scales and above, their abundance can constrain ULAs with masses of order $\mathcal{O}(10^{-26}\mathrm{~eV})$ and below. The inverse relation between scale and ULA mass highlights the importance of galaxy groups in the sample to explore higher ULA mass regimes. Very light ULAs with masses of order $\mathcal{O}(10^{-33}\mathrm{~eV})$ mimic a dark energy component, as BECs on the scale of the universe add a global energy component to the universe. Ultralight axions with masses slightly above this threshold undergo a transition from an effective dark energy component to an effective dark matter component at some redshift where the size of the universe exceeds the de Broglie wavelength of the ULA condensate. For this regime, no models exist for halo collapse in the presence of ULAs. Hence, we are not able to probe this region and exclude ULA masses $\logm < -32.5$ from our analysis.

\begin{figure}[h!]
    \centering
    \includegraphics[width=0.5\textwidth]{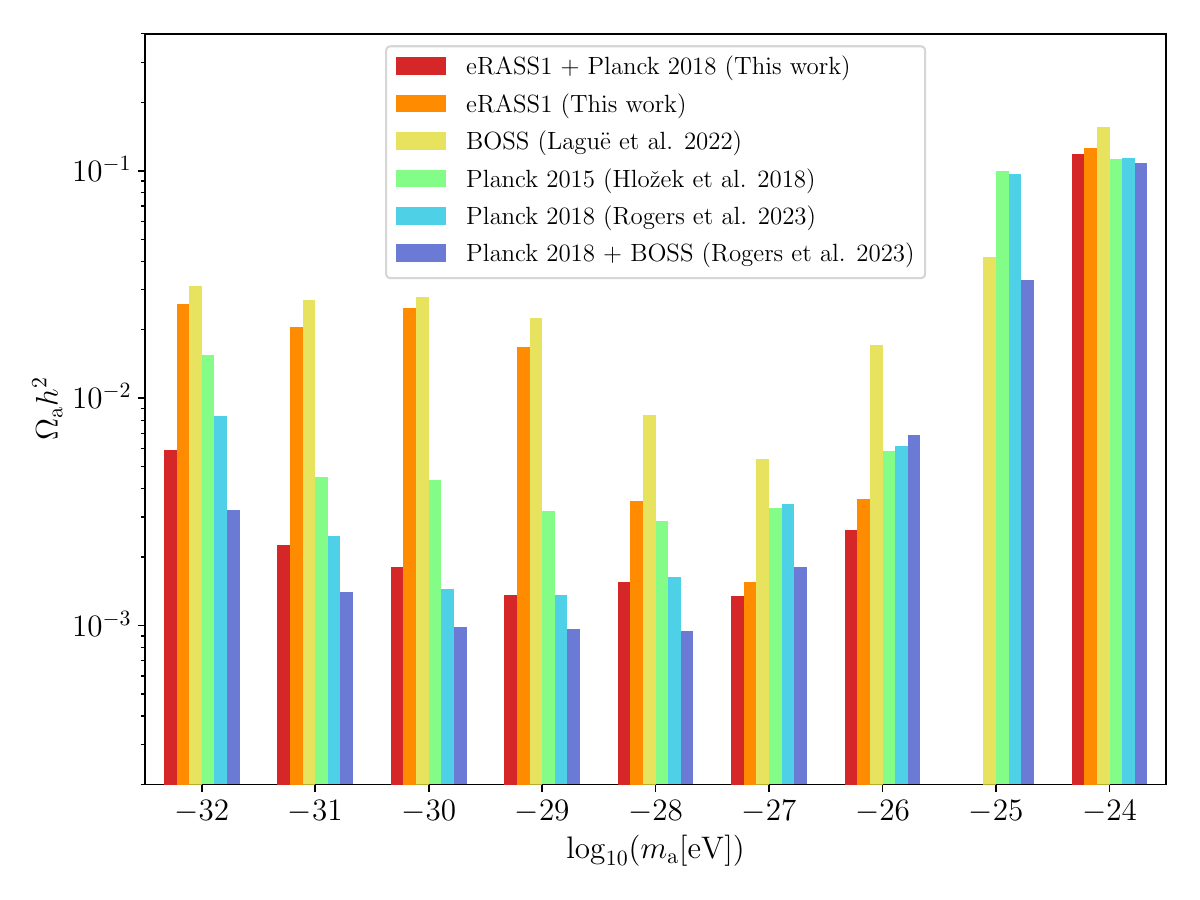}
    \caption{Comparison of ULA constraints obtained from eRASS1 galaxy cluster abundance (this work), \planck 2015 \citep{Hlozek2018}, and 2018 CMB data; BOSS galaxy clustering data \citep[both by][]{Rogers2023}; and combined analyses of eRASS1 and \planck 2015 (this work) and \planck 2018 and BOSS \citep{Rogers2023}. The bar height indicates the 95\% exclusion limits of the ULA density \oax times $h^2=(H_0/100)^2$ in each \logm bin. eRASS1 yields the tightest upper bounds on dark matter ultralight axions in the $\logm\in \{-27, -26\}$ bins.}
    \label{fig:constraints_comparison}
\end{figure}

We find upper bounds on the ULA relic density \oax obtained from binning the logarithmic ULA mass with a bin width of $\Delta\logm = 1$ around powers of ten. We treat each bin independently. We find upper bounds on the ULA relic density in each ULA mass bin with $\logm \leq -26$, as presented in Fig.~\ref{fig:constraints_2D}. We confirm the exclusion region below the $\logm=-25$ mass bin, which has been reported by \cite{Hlozek2015, Hlozek2018}, and \cite{Rogers2023}. eRASS1 galaxy cluster abundance alone yields the tightest constraints on the ULA relic density in the mass bins $\logm\in\{-27, -26\}$ with

\begin{align}
    \oax(-26.5<\logm\leq-25.5) &< 0.0079,\nonumber\\
    \oax(-27.5<\logm\leq-26.5) &< 0.0035,
\end{align}

\noindent both at a 95\% confidence level. Fig.~\ref{fig:constraints_comparison} shows that these are the tightest upper bounds that have been presented in the literature. Combining our results with the available chains from \planck 2015 CMB data \citep{Hlozek2018}, we find even tighter constraints in both mass bins:

\begin{align}
    \oax(-26.5<\logm\leq-25.5) &< 0.0058,\nonumber\\
    \oax(-27.5<\logm\leq-26.5) &< 0.0030.\\
\end{align}

From eRASS1 galaxy cluster number counts alone, we find that the ULA relic density \oax cannot exceed $2.7\%$ of the total energy density of the universe for $m_\mathrm{a} \leq 10^{-27.5}~\si{\electronvolt}$ and is bound below $0.84\%$ for ULA masses $10^{-27.5}~\si{\electronvolt} \leq m_\mathrm{a} \leq 10^{-26.5}~\si{\electronvolt}$ and below $0.36\%$ for ULA masses $10^{-26.5}~\si{\electronvolt} \leq m_\mathrm{a} \leq 10^{-25.5}~\si{\electronvolt}$. A significant contribution of ULAs to the total dark matter density can thus be ruled out by the growth of structure as observed through eRASS1 galaxy cluster number counts.

It is particularly interesting to compare the eRASS1 cluster number count constraints with those found using BOSS galaxy clustering data as another large-scale structure probe. \cite{Rogers2023} studied the cosmological implications of ULAs using the twelfth data release of the Baryon Oscillation Spectroscopic Survey (BOSS) catalog \citep{Dawson2013, Alam2017} and found upper bounds on the ULA fraction in a binned analysis in the range $-32 \leq \logm \leq -26$. As expected, galaxy clustering and galaxy cluster number counts constrain different ULA mass regimes. This is due to the relation between the ULA mass and the scale it affects: The heavier the ULA, the larger the affected scale in the matter power spectrum, for instance, a ULA with a mass of $\sim10^{-24}~\si{\electronvolt}$ is only observable on scales below $\sim200~\si{\kilo\parsec}$, while a ULA with a mass of $\sim10^{-25}~\si{\electronvolt}$ is already observable on scales $\sim500~\si{\kilo\parsec}$ and below (see Fig.~\ref{fig:matterpowerspectrum}). While, in comparison to other probes, \cite{Rogers2023} find the tightest constraints ($\oax h^2 < 0.0418$) in the ULA mass bin around $m_\mathrm{a} = 10^{-25}~\si{\electronvolt}$, corresponding to an effect on smaller scales, eRASS1 cluster number counts yield the tightest constraints among complementary in the mass bins $m_\mathrm{a} \in \{10^{-27}~\si{\electronvolt}, 10^{-26}~\si{\electronvolt}\}$. Ultralight axions in this mass range affect larger scales, which is constrainable using galaxy cluster number counts. Similar to galaxy clustering data, eRASS1 cluster number counts have reduced constraining power at scales extending the typical probe scales, as seen in the lowest-mass bins. On these largest scales (corresponding to  ULA masses $m_\mathrm{a} \lesssim 10^{-28}~\si{\electronvolt}$), CMB measurements by the \planck collaboration \citep{Planck2016b, Planck2020} have the highest constraining power and yield the tightest constraints \citep{Hlozek2015, Hlozek2018, Rogers2023}. For better comparison with the results obtained using other probes (see Fig.~\ref{fig:constraints_comparison}), we also provide the bounds of each constraining bin for \oax and $\oax h^2$ in Table~\ref{tab:mergedbounds}.

We emphasize that the weak lensing mass calibration as described in Sect.~\ref{sec:weak_lensing_calibration} remains valid over the full considered ULA mass range as the 95\% confidence levels do not exceed ULA fractions of $\oax/(\Omega_\mathrm{m}+\oax)\approx 0.1$ for $m_\mathrm{a} \lesssim 10^{-26}$~eV, and below $m_\mathrm{a} \approx 10^{-28.5}\,\si{\electronvolt}$, ULAs are not included in the HMF model and thus the modeling used in \citetalias{Ghirardini2024} remains valid. For higher ULA masses, Sect.~\ref{sec:weak_lensing_calibration} justifies the modeling choice of keeping NFW profiles for the weak lensing mass calibration up to any ULA fraction. 

Our current pipeline is limited by the model of the halo abundance in ULA cosmology, where we use a modified power spectrum in the \cite{Tinker2008} fitting function. Although this is justified by numerical simulations (see Sect.~\ref{sec:integration_of_ulas}), our results will be improved if suitable emulators are developed for this model. 
Additionally, for this pioneering work, we fix the masses of the neutrinos, while the effect of neutrinos is degenerate with ULAs. 
This modeling choice results in our constraints being conservative upper limits. Including massive neutrinos in the next analyses significantly tightened our constraints.

\subsection{Forecasts for the deepest eROSITA survey}
\label{sec:forecasts}

This section presents an estimation quantifying the potential of future galaxy cluster samples, illustrated by the deepest eROSITA All-Sky Surveys. The constraints presented in this work are based on eRASS1, which represents $\sim22$\% of the stacked consequent eROSITA All-Sky Survey (\erass[5], hereafter).
In the near future, the release of deeper eROSITA data and associated catalogs will significantly increase the number of available galaxy clusters and groups. In particular, we expect the number of low-mass galaxy groups to increase significantly with higher exposure time. To assess the expected evolution of cluster abundance in future eROSITA data releases, we generate a mock sample representing a catalog for \erass[5]. This mock catalog is constructed using the methodology outlined in \citetalias{Ghirardini2024} to validate the pipeline.

\begin{figure}[t]
    \centering
    \includegraphics[width=\linewidth]{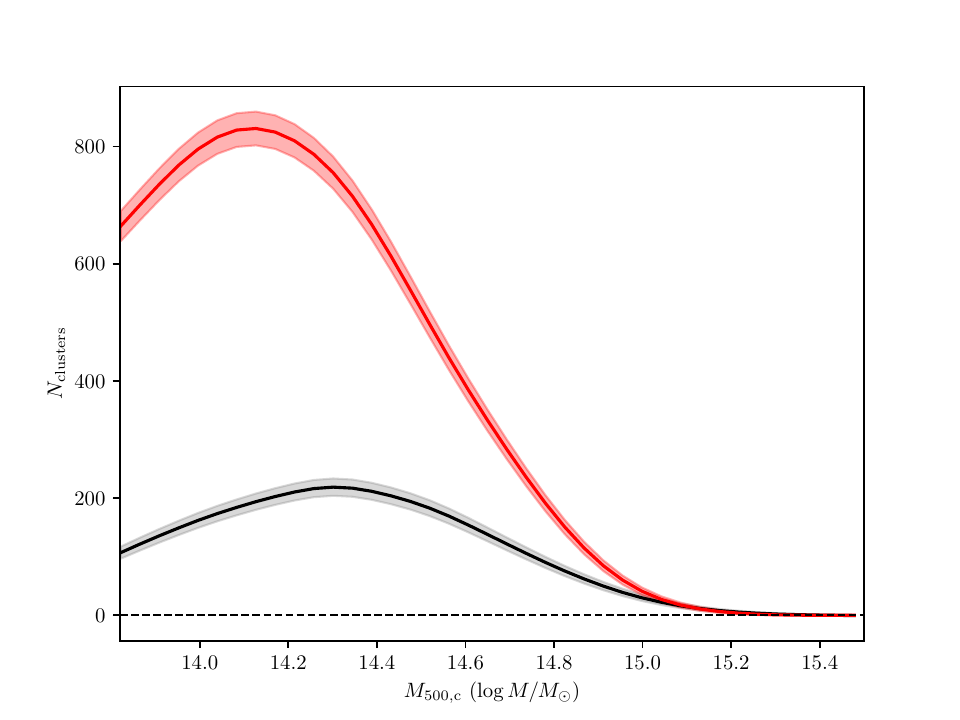}
    \includegraphics[width=\linewidth]{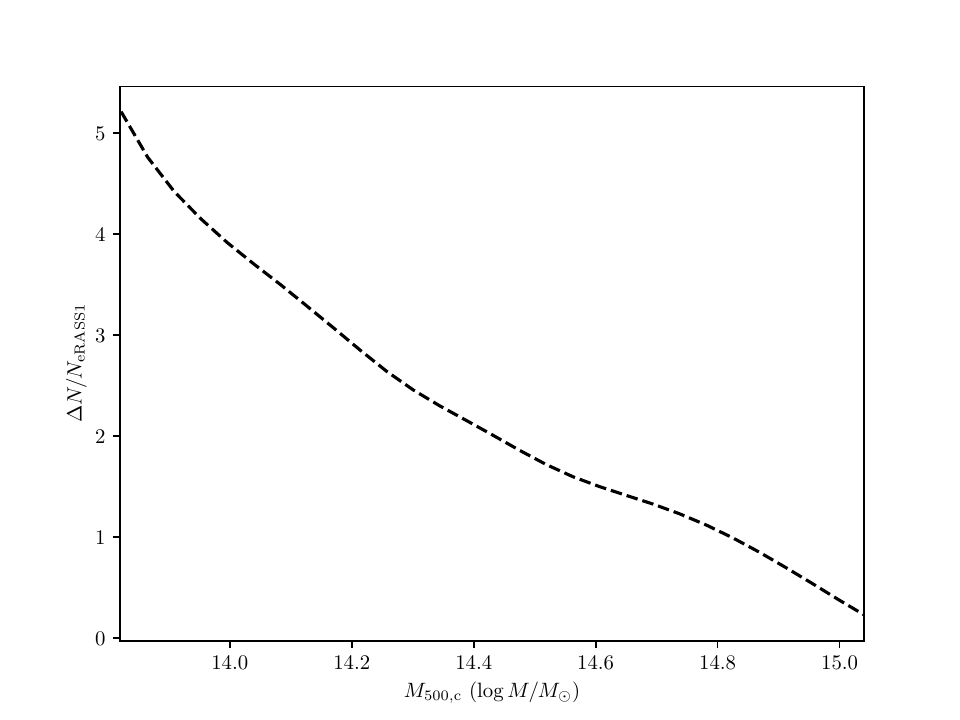}
    \caption{Forecasts for an \erass[5] survey. Left: Number count of X-ray detected clusters for an \erass[5] survey (red) compared with an eRASS1 catalog (black). The Poisson error bars are represented by filled lines. Right: Relative difference of \erass[5] detected clusters and eRASS1 detected objects. \erass[5] will contain five times more clusters at $M=10^{13.8}M_\odot$}.
    \label{fig:forecasts}
\end{figure}

We generated a dark matter halo catalog with the cosmology inferred in \citetalias{Ghirardini2024}, using the halo mass function from \cite{Tinker2008}. We then create mock X-ray and optical observables following the scaling relations described in section \ref{sec:stat_inf}. We then select the fraction of objects that would be detected using the selection function described in \cite{Clerc2024}. The X-ray selection function depends on X-ray count rates, hydrogen column density, background, and exposure time. We scale the exposure time by a factor of $4.6$ to simulate a survey with $4.6$ times the depth of eRASS1. The result is a catalog containing $\sim 15,000$ clusters with $0.1<z<0.8$. Figure \ref{fig:forecasts} presents the cluster abundance in eRASS1 and in the mock \erass[5] catalog. In the deeper eROSITA survey, the statistical power of cosmological constraints will be driven by the detection of a large and significant sample of galaxy groups.
At $10^{13.8}, M_\odot$, we detect approximately five times more massive clusters than in eRASS1, where the constraints on the parameter space are mostly constrained by the number of lower-mass galaxy clusters and galaxy groups (see Fig.~\ref{fig:hmf}).

\begin{figure}[h!]
\centering
\includegraphics[width=0.5\textwidth]{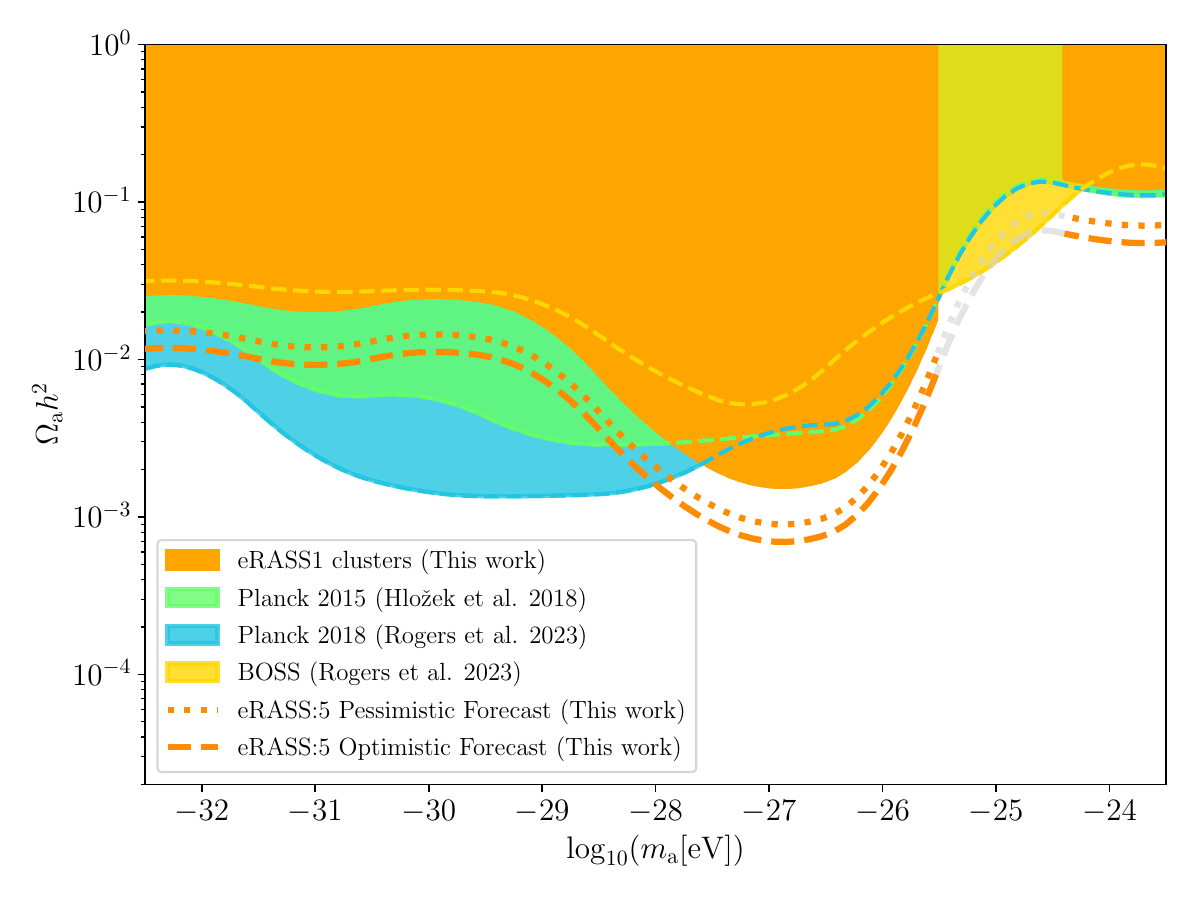}
\caption{Forecast for \erass[5] data. The exclusion region of the ULA parameter space obtained from \erass[5] mock catalog is presented with the orange dotted line for a pessimistic scenario and with the orange dashed line for an optimistic scenario, and can be compared with the plain orange region corresponding to the exclusion region obtained by eRASS1. For the missing bin, the \planck upper confidence level was assumed for the interpolation (see Sect.~\ref{sec:cosmo_constraints}).}
\label{fig:forecats_omega_a}
\end{figure}

We assumed that the Poisson error of the cluster abundance is the main statistical limiting factor. This means that we assume that the additional low-mass objects possess X-ray, optical, and weak lensing quantities measured as precisely as the ones obtained for massive clusters. We provide an optimistic and a pessimistic scenario based on the potential inclusion of low-mass clusters in Fig.~\ref{fig:forecats_omega_a}. In the optimistic case, our lower limit decreases by a factor of $2.2$, while it is lowered by a factor of $1.7$ in the pessimistic case. The reliability of our forecasts for the \erass[5] data is influenced by the accuracy of the selection function; the robustness of the model used to predict the abundance, including the HMF model; the universality of the scaling relations over the full survey mass range; and the precision of forthcoming weak lensing measurements for the mass calibration of low-mass galaxy groups.

\section{Conclusions}
\label{sec:conclusion}

The unique properties of ULAs lead to observable phenomena such as solitonic cores in dark matter halos and interference patterns on cosmological scales. These features can be probed through astrophysical observations, providing potential evidence of their existence.

Galaxy cluster number counts inferred from the eRASS1 cluster cosmology catalog \citep{Bulbul2024, Kluge2024} can be used to test and constrain the viable dark matter candidates, such as ULAs, through the evolution of the structure growth. Indeed, eRASS1, being the largest ICM-selected sample to date, has a significant constraining power. Additionally, it has the potential to probe the group regime, which is very sensitive to dark matter signatures. 

In this work, we performed Bayesian inference on cosmological parameters in a universe with an additional ULA component parameterized by the logarithmic ULA mass and ULA relic density $\{\logm, \oax\}$. The matter power spectrum was computed by \texttt{axionCAMB} \citep{Hlozek2015}, from which we computed the HMF using the multiplicity function by \cite{Tinker2008}. The forward modeling process requires the selection function developed by \cite{Clerc2024} and additional parameters from the mixture model that take care of possible contaminants, the count-rate-mass scaling relation, and the richness-mass scaling relation, which are calibrated by gravitational weak lensing shear maps \citep{Grandis2024}. Apart from the adaptations described in Sect.~\ref{sec:integration_of_ulas}, we used the framework established in \citetalias{Ghirardini2024} for the parameter inference.

We performed a binned analysis with logarithmic ULA mass bins covering the whole mass spectrum of ultralight axions from the dark matter regime starting at $\logm=-24$ down to the dark energy regime with the lowest ULA mass at $\logm=-32$ including an independent full cosmological inference in each bin while assuming a flat universe. We constrained the following set of cosmological parameters: the matter relic density, $\Omega_{\rm m}$, and the normalization of the matter power spectrum, $\log(A_{\rm s})$. We left the Hubble constant, $H_0$; the scalar spectral index, $n_{\rm s}$; and the baryon relic density, $\Omega_{\rm b}$, free but could constrain them as done in \citetalias{Ghirardini2024}. We chose a uniform prior on \oax between 0 and 1 in both scenarios. All common parameters have the same priors as in \citetalias{Ghirardini2024}. 

We performed a consistency check with the eRASS1 cosmological results from \citetalias{Ghirardini2024} by fixing ULA parameter values to $\logm \gg -18$ and $\oax \rightarrow 0$, effectively switching off the effect of ULAs on structure growth. The results are consistent with \citetalias{Ghirardini2024}, indicating that including ULAs does preserve the correctness and reliability of the inference framework.

The binned analysis of eRASS1 galaxy cluster number counts alone yields tight bounds on the ULA relic density, especially in the two highest ULA mass bins within the exclusion region $\logm \in \{-27, -26\}$. This ULA mass range corresponds to physical scales (de Broglie wavelengths) of ULA BECs comparable to the typical galaxy cluster scales around $1$~Mpc to $10$~Mpc. The parameter space excluded by eRASS1 galaxy clusters in comparison to complementary probes is shown in Fig.~\ref{fig:exclusionregion}. In the mass range of $10^{-27.5}~\si{\electronvolt} \leq m_\mathrm{a} \leq 10^{-26.5}~\si{\electronvolt}$, the upper limit on the ULA relic density is $\oax < 0.0036$ of the total energy density of the universe. Similarly, in the mass range of $10^{-26.5}~\si{\electronvolt} \leq m_\mathrm{a} \leq 10^{-25.5}~\si{\electronvolt}$, the ULA relic density is constrained to $\oax < 0.0084$ of the total energy density of the universe. The $m_\mathrm{a}=10^{-25}$~eV mass bin was excluded from the analysis due to a lack of modeling and statistical robustness in that particular bin. In the mass regime $10^{-32.5}~\si{\electronvolt} \leq m_\mathrm{a} \leq 10^{-27.5}~\si{\electronvolt}$, the ULA relic density is bound below $\oax \lesssim 0.055$. We conclude that the evolution of structure growth as observed by eRASS1 galaxy cluster number counts rules out ULAs as a significant contribution to the total dark matter density of the Universe. The constraints on the abundance at different confidence levels concerning a logarithmic ULA mass bin are presented in Fig.~\ref{fig:constraints_2D}. Overall, these results highlight the potential of cluster abundance measurements and the evolution of the growth of structure as a probe of ULAs. For the first time, we have obtained constraints using cluster abundance in a redshift and scale regime complementary to the one probed by other cosmological probes.

In the near future, statistically more powerful cosmological simulations, including ULAs (or ultralight scalar fields, fuzzy dark matter) that predict an HMF model and shear profiles for a sufficiently large cosmological parameter space, will be required to provide unbiased constraints. 
We note that the constraints presented in this work were obtained using the eRASS1 cosmology sample of 5,259 X-ray-selected massive clusters. The primary sample contains a higher fraction of low-mass objects, so we are more sensitive to ULA signatures, provided that we carefully model the baryonic systematic effects that affect groups. Future analyses of deeper surveys with \erosita at \erass[5] depth will provide larger samples of low-mass halos with a greater constraining power in the ULA parameter space, improving the overall constraints by a factor of $1.7$.

\begin{acknowledgement}

The authors thank the referee for the helpful and constructive
comments on the draft. We further thank Christian Garrel for his contributions to the codebase and the valuable exchange with him. Thanks to Johannes Diehl, Julia Sisk-Reynes, and Luis A. Ure\~na-L\'opez for the helpful discussions.

This work is based on data from eROSITA, the soft X-ray instrument aboard SRG, a joint Russian-German science mission supported by the Russian Space Agency (Roskosmos), in the interests of the Russian Academy of Sciences represented by its Space Research Institute (IKI), and the Deutsches Zentrum f{\"{u}}r Luft und Raumfahrt (DLR). The SRG spacecraft was built by Lavochkin Association (NPOL) and its subcontractors and is operated by NPOL with support from the Max Planck Institute for Extraterrestrial Physics (MPE).

The development and construction of the eROSITA X-ray instrument was led by MPE, with contributions from the Dr. Karl Remeis Observatory Bamberg \& ECAP (FAU Erlangen-Nuernberg), the University of Hamburg Observatory, the Leibniz Institute for Astrophysics Potsdam (AIP), and the Institute for Astronomy and Astrophysics of the University of T{\"{u}}bingen, with the support of DLR and the Max Planck Society. The Argelander Institute for Astronomy of the University of Bonn and the Ludwig Maximilians Universit{\"{a}}t Munich also participated in the science preparation for eROSITA.

The eROSITA data shown here were processed using the \esass software system developed by the German eROSITA consortium.

S. Zelmer, E. Bulbul, V. Ghirardini, A. Liu, and X. Zhang acknowledge financial support from the European Research Council (ERC) Consolidator Grant under the European Union’s Horizon 2020 research and innovation program (grant agreement CoG DarkQuest No 101002585). N. Clerc was financially supported by CNES. T. Schrabback and F. Kleinebreil acknowledge support from the German Federal Ministry for Economic Affairs and Energy (BMWi) provided through DLR under projects 50OR2002, 50OR2106, and 50OR2302, as well as the support provided by the Deutsche Forschungsgemeinschaft (DFG, German Research Foundation) under grant 415537506. M. Br\"uggen acknowledges funding by the Deutsche Forschungsgemeinschaft (DFG, German Research Foundation) under Germany’s Excellence Strategy – EXC 2121 “Quantum Universe” – 390833306. S. Krippendorf’s work has been
partially supported by STFC consolidated grants ST/T000694/1 and ST/X000664/1.
\end{acknowledgement}

\bibliography{ULAreferences}

\end{document}